\documentclass[preprint,pteplogo]{ptephy}
\usepackage{float}   
\preprintnumber{1306.6683} 

\usepackage{amsmath} 
\usepackage{amsthm} 
\usepackage{hyperref} 
\usepackage{graphics} 
\usepackage{algorithmic} 
\usepackage{subfig} 
\usepackage{url} 

\hypersetup{
    bookmarks=true,         
    unicode=false,          
    pdftoolbar=true,        
    pdfmenubar=true,        
    pdffitwindow=true,     
    pdfstartview={FitH},    
    pdftitle={My title},    
    pdfauthor={Author},     
    pdfsubject={Subject},   
    pdfcreator={Creator},   
    pdfproducer={Producer}, 
    pdfkeywords={keyword1} {key2} {key3}, 
    pdfnewwindow=true,      
    colorlinks=true,       
    linkcolor=red,          
    citecolor=red,        
    filecolor=magenta,      
    urlcolor=cyan           
}

\newcommand{\bk}{{\bf k}}
\newcommand{\bp}{{\bf p}}
\newcommand{\bq}{{\bf q}}

\begin{document}

\title{Three-Body Bound State Calculations by Using Three-Dimensional Low Momentum Interaction $V_{low-k}$}

\author{\name{M. R. Hadizadeh}{1,\ast}\thanks{The calculations of this paper is done during the author's appointment to Instituto de F\'\i sica Te\'orica, Universidade Estadual Paulista, 01140-070, S\~ao Paulo, SP, Brazil.}}

\address{\affil{1}{Institute of Nuclear and Particle Physics, Department of Physics and Astronomy, Ohio University, Athens, OH 45701, USA}
\email{hadizadm@ohio.edu}}

\begin{abstract}%
Three-dimensional (3D) Faddeev integral equations are solved for three-body (3B) bound state problem without using the partial wave (PW) form of low momentum two-body (2B) interaction $V_{low-k}$ which is constructed from spin independent Malfliet-Tjon V (MT-V) potential. The dependence of 3B binding energy on the cutoff momentum of $V_{low-k}$ is investigated for a wide range of $\Lambda$ from $1.0$ to $7.0\, fm^{-1}$. The properties of Faddeev components and 3B wave function are displayed and the effect of number of grid points for momentum and angle variables on the accuracy and the stability of numerical results is studied by calculation of the expectation value of total Hamiltonian.
\end{abstract}

\subjectindex{D05, D00, D06}

\maketitle

\section{Introduction}\label{intro}
The low-momentum nucleon-nucleon ($NN$) interaction $V_{low-k}$ is derived in PW representation, based on the strategy of integrating out the high momentum part of $NN$ interaction $V_{NN}$. 
The properties of $NN$ system for laboratory energies $E_{lab} < 350$ MeV (like deuteron binding energy, low-energy phase shifts and half-on-shell $T-$matrix of bare potential $V_{NN}$) are all preserved by $V_{low-k}$ which is derived from very different $NN$ potential models (like Paris, CD Bonn, Argonne, Idaho and the Nijmegen potentials), when it is confined within a cutoff momentum $\Lambda \sim 2.0\,fm^{-1}$.
Different methods are developed in PW representation to derive the $V_{low-k}$, such as 
Renormalization Group (RG) \cite{Bogner-NPA684}-\cite{Bogner-PLB576} and the model space techniques like Lee-Suzuki similarity transformations \cite{Lee-PLB91,Suzuki-PTP92}. 
It is shown that the Lee-Suzuki method is equivalent to Renormalization Group (RG) method and reproduces the same results, whereas its numerical procedure is less cumbersome in comparison to solution of differential equation in RG method. 

On this basis, recently the three-dimensional (3D) form of low-momentum $NN$ interaction $V_{low-k}$ is formulated, into the model space Lee-Suzuki method, as a function of momentum vectors and is calculated for a spin-independent potential model \cite{Bayegan-NPA814} and also for a modern $NN$ interaction \cite{Bayegan-NPA832}. It is shown that similar to PW representation, the 3D form of low momentum interaction reproduce the same $NN$ observables from bare potential $V_{NN}$, whereas it avoids the highly involved angular momentum algebra occurring for transition and permutation operators of PW representation.

In this paper we have employed the spin-independent $V_{low-k}$ constructed from MT-V potential \cite{Malfliet-NPA127}, which acts on all PWs, to calculate 3B binding energy in a non PW scheme by solution of 3D Faddeev integral equations. The momentum cutoff dependence of binding energy is studied and it is shown that the 3B binding energy is strongly cutoff dependent at low cutoff values, similar to what is predicted by PW-based calculations \cite{Fujii-PRC70,Nogga-PRC70}. 

Implementation of the 3D form of low momentum $NN$ interaction obtained with the similarity renormalization group (SRG) method \cite{Bogner-PRC75} in $NN$ scattering \cite{Bayegan-PRC79} as well as three- and four-body bound states calculations \cite{Bayegan-PRC77}-\cite{Hadizadeh-PRC83} are in progress.

The paper is organized as follows. In Sect. \ref{Sec.Vlowk} the 3D formalism of $V_{low-k}$ for spin-independent case is briefly presented and our numerical results for matrix elements of $V_{low-k}$ are given. In Sec. \ref{Faddeev-integral-equation} we review the three-dimensional Faddeev integral equations for 3B bound state and in Sec. \ref{Numerical-results} our numerical results for 3B binding energies and momentum distribution functions for low momentum potentials, and for a wide range of cutoff momentums are given. To test the accuracy of our numerical results the expectation value of 3B Hamiltonian is calculated and compared with eigenvalue binding energy. We conclude in Sec. \ref{Summary}.

\section{3D representation of $V_{low-k}$ in model space Lee-Suzuki method}
\label{Sec.Vlowk}

The low-momentum interaction in the model space Lee-Suzuki, which reproduces the model space
components of the wave function from the full-space wave function, 
is given by:
\begin{eqnarray} \label{eq:Vlow-operator}
V_{low-k}=PV_{NN}(P+Q \omega P),
\end{eqnarray}
where $V_{NN}$ is bare 2B interaction, $P$ and $Q$ are 2B projection operators which project a state onto the model space (low-momentum space) and its complement (high-momentum space), respectively. $\omega$ is an operator which transforms the states of the $P$ space to the states of the $Q$ space. As shown in Ref. \cite{Bayegan-NPA814}, the momentum space representation of Eq. (\ref{eq:Vlow-operator}) by considering the integral form of the projection operators $P$ and
$Q$, and in 3D approach reads:
\begin{eqnarray} \label{eq.Vlowk}
V_{low-k}(\bp',\bp)=V_{NN}(\bp',\bp)+\int_{\Lambda \le k <\infty}
d^3\bk \,\,V_{NN}(\bp',\bk)\,\omega(\bk,\bp),
\end{eqnarray}
where $\bp$, $\bp'$ are 2B momentum vectors in model space $P$ and $\bk$ is the 2B momentum in the complement model
space $Q$. 
The matrix elements of $\omega(\bk,\bp)$ can be obtained by solution of following integral equation:
\begin{eqnarray} \label{eq.omega}
\omega(\bk,\bp)=\int_{0 \le p' \le \Lambda}d^3p'\,\,\,\Psi_{\bp'}^{NN}(\bk)\, \tilde{\Psi}_{\bp'}^{NN}(\bp),
\end{eqnarray}
where $\Psi_{\bp'}^{NN}(\bp)$ and
$\Psi_{\bp'}^{NN}(\bk)$ are the wave function
components of the $P$ and $Q$ spaces of the full-space
respectively. They are given in the form of the half-on-shell (HOS) 2B
$T$-matrix by:
\begin{eqnarray}
\Psi_{\bp'}^{NN}(\bk)&=&
\frac{T(\bk,\bp',k^{2})}{\dfrac{p'^2}{m}-\dfrac{k^2}{m}},
 \\
\Psi_{\bp'}^{NN}(\bp)&=&\delta^3(\bp-\bp')+
\frac{T(\bp,\bp',p'^{2})}{\dfrac{p'^{2}}{m}-\dfrac{p^{2}}{m}+i\varepsilon},
\end{eqnarray}
where the HOS 2B $T$-matrix
can be obtained from the Lippmann-Schwinger equation in the 3D
representation \cite{Elster-FBS24}:
\begin{eqnarray} \label{eq.T-matrix}
T(\bp',\bp,p^{2})=V_{NN}(\bp',\bp)+
\int\emph{d}^{3}p''
\frac{V_{NN}(\bp',\bp'')\,T(\bp'',\bp,p^{2})}{\dfrac{p^{2}}{m}-\dfrac{p''^{2}}{m}+i\varepsilon}.
\end{eqnarray}
The $V_{low-k}$ given by the Lee-Suzuki method is non-Hermitian, i.e. $V_{low-k}(\bp',\bp) \ne V_{low-k}(\bp,\bp')$, and specifically constructed to preserve the HOS $T-$matrix $T(\bp',\bp,p^{2})$. This interaction of course preserves the phase shift which is given by the fully-on-shell $T-$matrix $T(\bp,\bp,p^{2})$.
\begin{figure}[H]
\begin{center}
\includegraphics[width=5.5in]{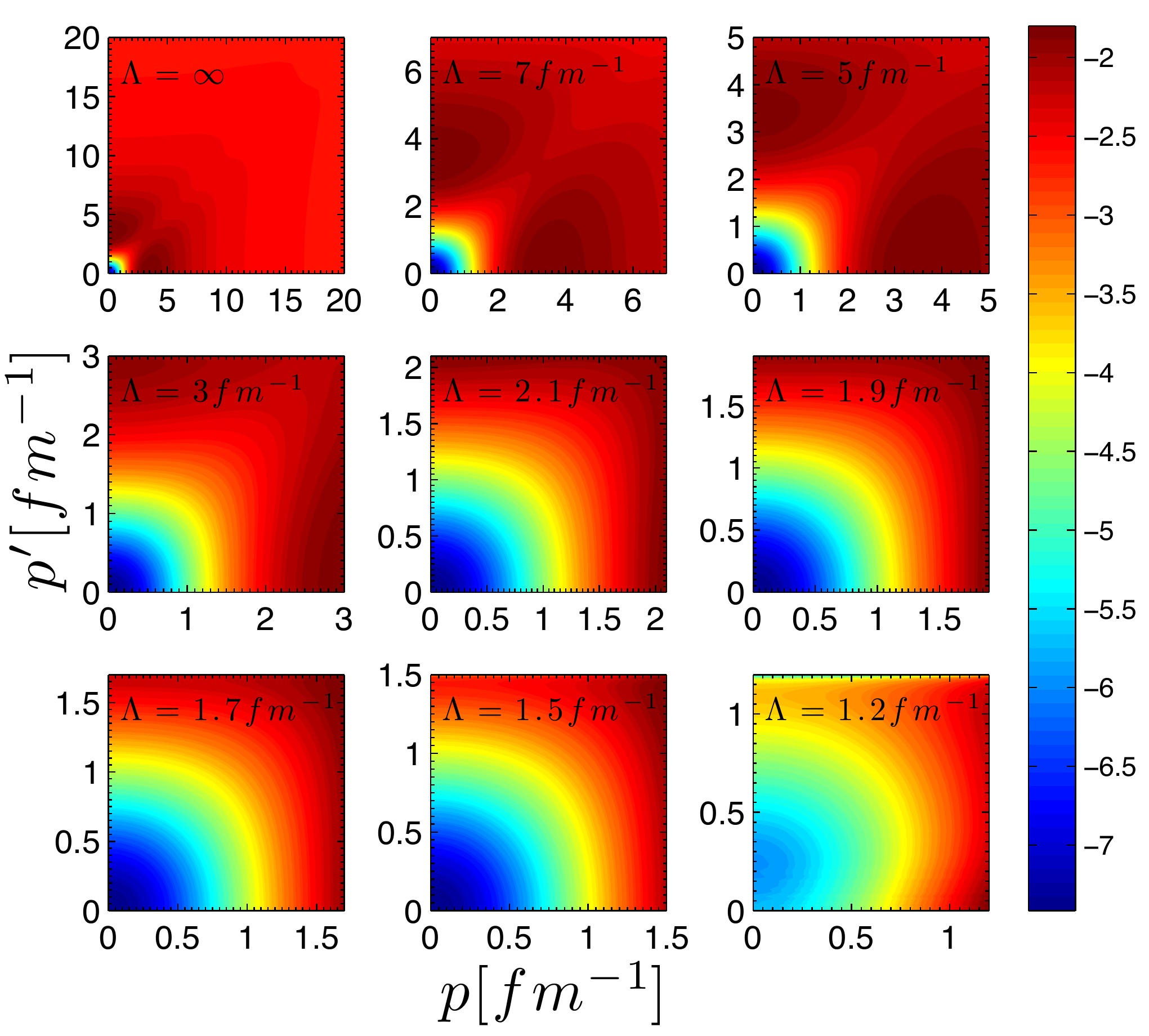} 
\end{center}
\caption{Angle averaged bare and constructed low momentum interactions calculated from MT-V potential as a function of momentum variables $p$ and $p'$.}
\label{fig:MT-V-pp-average}
\end{figure}
Details of numerical solution of 3D integral equations (\ref{eq.Vlowk}), (\ref{eq.omega}) and (\ref{eq.T-matrix}) by choosing suitable coordinate systems are given in Ref. \cite{Bayegan-NPA814} and we don't repeat it here. As it is shown in this reference if we choose the Jacobi momentum vector $\bp$ parallel to $z$
axis and the vector $\bp'$ is in $x-z$ plane, we can obtain the matrix elements of low momentum interaction $V_{low-k}(p',p,x')$ from solution of integral equation (\ref{eq.Vlowk}), where $p$ and $p'$ are the magnitude of Jacobi momenta and $x'$ is the angle between them. 

In this paper for numerical solution of integral equations we have used spin independent MT-V potential with the parameters given in Ref. \cite{Elster-FBS27}. In Fig. \ref{fig:MT-V-pp-average} we have shown the 2D plots of the angle averaged low momentum interaction $V^{ave}_{low-k}(p',p)=\frac{1}{2}\int_{-1}^{+1} dx' \, V_{low-k}(p',p,x')$ as a function of momentum variables $p$ and $p'$ for a range of cutoffs $\Lambda$ from $1.2$ to $7.0\, fm^{-1}$. It indicates that the non-Hermiticity of $V_{low-k}(\bp',\bp)$ is rather weak and it is a smooth potential for cutoff momentum $\Lambda$ in the vicinity $2.0\,fm^{-1}$. Bare MT-V potential is also shown as $\Lambda=\infty$.  

\section{3D Faddeev integral equations for 3B bound state}
\label{Faddeev-integral-equation}
The bound state of three particles which interact by pairwise forces can be described by Faddeev equation:
\begin{equation}
|\psi\rangle \equiv |\psi_{12,3}\rangle =G_{0} \, t \,P \, |\psi\rangle,
\label{eq.Faddeev}
\end{equation}
where $G_0=(E-H_0)^{-1}$ is free propagator, $P = P_{12}P_{23}+P_{13}P_{23}$ is permutation operator. $t = V + V \, G_0 \, t$ is 2B $t-$matrix which can be obtained from Lippmann-Schwinger equation, where $V$ is applied to the low-momentum potential or the original one. 
Total wave function of 3B system can be obtained from Faddeev component (\ref{eq.Faddeev}) as:
\begin{equation}
|\Psi\rangle = (1+P) \, |\psi\rangle.
\label{eq.WF}
\end{equation}
In order to project Eq. (\ref{eq.Faddeev}) in momentum space, we need to define 3B basis states which are composed of two Jacobi momentum vectors $\bp$ and $\bq$. The momentum vector $\bp=\frac{\bk_2-\bk_3}{2}$ is the relative momentum of the pair in 2B subsystem and $\bq=\frac{2}{3} (\bk_1 -\frac{1}{2}(\bk_2+\bk_3) )$ is the relative momentum of third particle to the center of mass of pair. 3B basis states $|\, \bp \, \bq \,\rangle$ are complete and normalized as:
\begin{equation}
 \int d^{3}p \, \int d^{3}q \,\,|\, \bp \, \bq
\,\rangle \,\langle\, \bp \, \bq \,  |=\mathbf{1}, \quad \langle \, \bp \, \bq\, |\,\bp' \, \bq' \, \rangle=\delta^{3}({\bf
p}-\bp') \, \delta^{3}(\bq-\bq').
\label{eq.normalization}
\end{equation}
The projection of Faddeev component, Eq. (\ref{eq.Faddeev}), in introduced basis states of Eq. (\ref{eq.normalization}) reads:
\begin{equation}
\psi (\bp , \bq) = \frac{1}{ E- \frac{p^2}{m} -
 \frac{3q^2}{4m} } \, \int d^3q' \; t_{s} \bigl (\bp,\frac{1}{2}
 \bq+\bq';  \epsilon   \bigr) \,   \psi \bigl ( \bq+
 \frac{1}{2}\bq',\bq' \bigr ),  
 \label{eq.Fc-momentum}
\end{equation}
where the two-body subsystem energy is defined as $\epsilon=E-\dfrac{3q^2}{4m}$. The symmetrized 2B $t-$matrix in the kernel of integral equation is $t_{s}(\bp,\bq;E)=t(\bp,\bq;E)+t(-\bp,\bq;E)$. 
Total 3B wave function can be obtained as:
\begin{equation}
 \Psi (\bp \, \bq) =\psi (\bp \, \bq) + \psi \bigl (-\frac{1}{2}\bp-\frac{3}{4} \bq, \bp-\frac{1}{2}\bq \bigr)+ \psi \bigl (-\frac{1}{2}\bp+\frac{3}{4} \bq, -\bp-\frac{1}{2}\bq \bigr).
 \label{eq.WF-momentum}
\end{equation}
Since we have ignored spin-isospin degrees of freedom and we study three-boson bound state, the Faddeev amplitude $\psi (\bp , \bq)$ and consequently total wave function $\Psi (\bp , \bq)$ should be symmetric under the exchange of interacting particles in 2B subsystem. On the other hand they are symmetric under exchange $\bp$ to $-\bp$ which can be easily verified from Eqs. (\ref{eq.Fc-momentum}) and (\ref{eq.WF-momentum}).
In order to numerically solve the integral equation (\ref{eq.Fc-momentum}), as shown in Ref. \cite{Elster-FBS27}, choosing a suitable coordinate system, $\bq$ parallel to $z-$axis and $\bp$ in $x-z$ plane, leads to a three-dimensional integral equation:
\begin{equation}
\psi(p,q,x) = \frac{1}{{E-\frac{p^{2}}{m}
-\frac{3q^{2}}{4m}}} \,  \int_{0}^{\infty} dq' \, q'^{2}
\int_{-1}^{+1} dx' \int_{0}^{2\pi} d\varphi'   \, t_{s} \biggl ( p,\tilde{\pi}, x_{p \tilde{\pi}} ;\epsilon \biggr) \, \psi(\pi, q', x_{\pi q'}),
 \label{eq.FC-magnitude}
 \end{equation}
where the angle variables and shifted momentum arguments are given as:
\begin{eqnarray}
x_{qp} &\equiv& \hat{\bq}.\hat{\bp} = x \nonumber \\*
x_{qq'} &\equiv& \hat{\bq}.\hat{\bq}'  = x' \nonumber \\*
x_{pq'}&\equiv& \hat{\bp}.\hat{\bq}' = xx'+\sqrt{1-x^{2}}\sqrt{1-x'^{2}}\cos (\varphi')
 \nonumber \\*
\tilde{\pi}&=&\sqrt{\frac{1}{4}q^{2}+q'^{2}+qq'x'} 
 \nonumber \\*
 x_{p \tilde{\pi}}&=&\frac{\frac{1}{2}qx+q'x_{pq'}}{\tilde{\pi}}
  \nonumber \\* 
  \pi&=&\sqrt{q^{2}+\frac{1}{4}q'^{2}+qq'x'}
   \nonumber \\*
x_{\pi q'} &=& \frac{qx'+\frac{1}{2}q'}{\pi}  \label{eq.variables}
 \end{eqnarray}

By having the Faddeev amplitude $\psi(p,q,x)$, the three-body wave function $\Psi(p,q,x)$ can be obtained from Eq. (\ref{eq.WF-momentum}) and by considering the coordinate system defined in Eq. (\ref{eq.variables}).

\section{Numerical results}
\label{Numerical-results}

\subsection{3B binding energy and wave function}\label{3B-BE}

We have solved the 3D Faddeev integral equation (\ref{eq.Fc-momentum}) for bare and low momentum interactions constructed from MT-V potential. For numerical solution of this integral equation we have first discretized continues momentum and angle variables. To this aim we have used Gauss-Legendre quadrature with 40 mesh points for all momentum and angle variables. 
For discretization of momentum variables $p$ and $q$ we need to determine their corresponding momentum cutoffs $p^{max}$ and $q^{max}$. In order to avoid the extrapolation on symmetrized two-body $t-$matrix $t_{s} \bigl ( p,\tilde{\pi}, x_{p \tilde{\pi}} ;\epsilon \bigr)$ and Faddeev component $\psi(\pi, q', x_{\pi q'})$ for solution of integral equation (\ref{eq.FC-magnitude}), the following condition should be satisfied:
\begin{eqnarray}
\tilde{\pi}^{max}=\pi^{max} =  1.5\, q^{max} \le p^{max} ,
 \end{eqnarray}
whereas we have considered same momentum cutoff for $q$ and $q'$, i.e. $q^{max}$. 
A linear mapping $\frac{c}{2}.(1+x)$ is used for discretization of Jacobi momenta $p$ and $q$, where $x$ are the roots of Gauss-Legendre polynomial. In the calculations with bare potential, we have used $c=20.0\, fm^{-1}$ and $c=7.0\, fm^{-1}$ for Jacobi momentum $p$ and $q$, respectively. These cutoffs satisfy the condition $p^{max} \ge 1.5\, q^{max}$. 
Since the low momentum potential $V_{low-k}$, which is dependent to relative two-body momenta $p$ and $p'$, is confined in a low momentum space with a cutoff $\Lambda$, consequently the symmetrized two-body $t-$matrix, which appears in the kernel of Faddeev integral equation is also defined in a momentum interval from $0$ to $\Lambda$. 
So, for 3B calculations with low momentum potentials, $c=\Lambda$ is used for both $p$ and $q$ Jacobi momenta. Of course, for this case, in each step of iteration, we equate the value of Faddeev component $\psi(\pi, q', x_{\pi q'})$ for $\pi^{max}>\Lambda$ to zero.
 The integral equation is solved by Lanczos-type technique, which is based on iteration (see Appendix C2 of Ref. \cite{Hadizadeh-PRA85}).    
The iteration of the integral Eq. (\ref{eq.Fc-momentum}) requires a very large number of two-
dimensional interpolations on the Faddeev component and 2B $t-$matrix, to this aim we have used Cubic-Hermitian Splines to reach high computational accuracy and speed. Usually 7 to 10 iterations is enough to reach the convergence in the solution of integral equation. 
\begin{table}[H]
\centering
\caption {Three-body binding energy for bare and low momentum interaction for MT-V potential.}
\label{table:BE}
\begin{tabular}{c|ccccccccccccccccccccccl}
\hline \hline
 $\Lambda \, [fm^{-1}]$ & 1.0 &1.1 & 1.2 & 1.3 & 1.4 & 1.5  \\
 $E_3$ [MeV]  & -6.146  & -6.568 & -6.917 & -7.190 & -7.395 & -7.544 \\
 \hline 
 $\Lambda \, [fm^{-1}]$ &  1.6 & 1.7 & 1.8 & 1.9 & 2.0 & 2.1  \\
 $E_3$ [MeV]  & -7.646 & -7.713 & -7.754 & -7.777 & -7.788  & -7.792 \\
 \hline
  $\Lambda \, [fm^{-1}]$ & 2.5 & 3.0 & 4.0 & 5.0 & 6.0 & 7.0  \\
  $E_3$ [MeV]   & -7.783 & -7.766 & -7.744 & -7.736 &-7.737 & -7.738 \\
  \hline
  \hline
  $V_{NN}$ & -7.738  \\
\hline \hline
\end{tabular}
\end{table}

\begin{figure}[H]
\begin{center}
\includegraphics[width=5.0in]{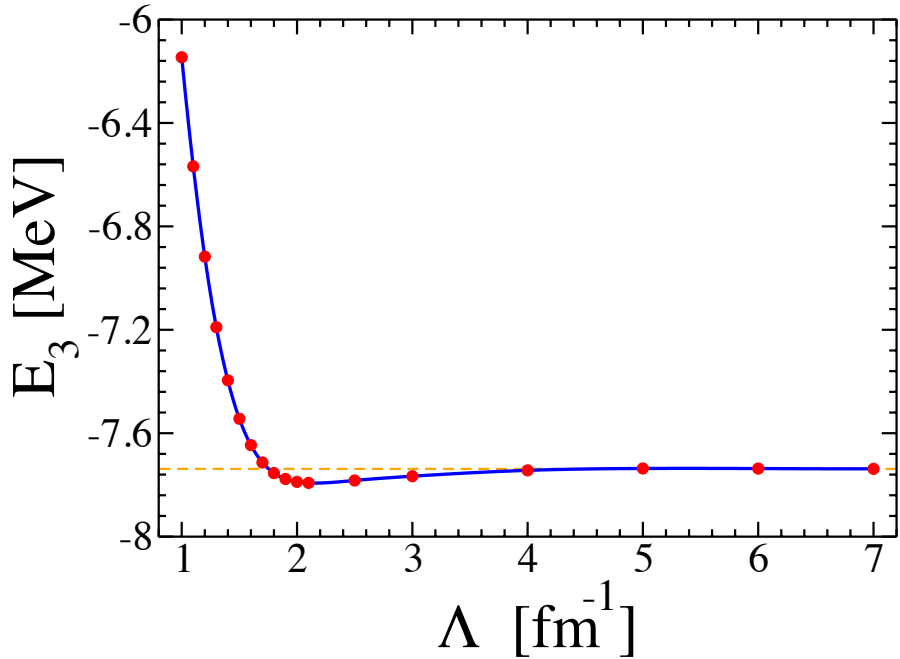} 
\end{center}
\caption{The dependence of three-body binding energy $E_3$ to the cutoff of low-momentum potential $\Lambda$.}
\label{fig:3B-Lambda}
\end{figure}

Our numerical results for 3B binding energy for a wide range of cutoffs $\Lambda$, from $1.0$ to $7.0\, fm^{-1}$ are given in Table \ref{table:BE}. The binding energies are obtained in such a way that Eq. (\ref{eq.Fc-momentum}) is fulﬁlled with a relative accuracy of $10^{-6}$ at each set points $(p,q,x)$. In Fig. \ref{fig:3B-Lambda}, we have shown the binding energy results as a function of cutoff $\Lambda$. 
Obviously the 3B binding energy is strongly cutoff dependent for small values of cutoffs, whereas for values larger than $5.0\, fm^{-1}$ it is almost cutoff independent and leads to bare potential binding energy $-7.74$ MeV. 
This cutoff dependency is quite reasonable, because 2B interaction has a cutoff $\Lambda$ (P-space of QCD) and therefore has corresponding three- and higher-body forces. Consequently, if we omit the many-body forces, the observables will be cutoff-dependent. By using $V_{low-k}$ all low-energy 2B observables are cutoff-independent, and therefore, by varying the cutoff $\Lambda$ in our 3B calculations we can evaluate the effects of the omitted 3B forces.
Our results show that the cutoff $\Lambda$ variations from $1.0$ to $7.0 \, fm^{-1}$ leads to approximately $1.6$ MeV variations for 3B binding energy, which gives an estimate of the 3B forces contribution. 
In order to reproduce the bare solution and achieve the cutoff independent results for low momentum cutoffs, we need to consider the 3B forces induced by a unitary transformation of the Hamiltonian \cite{Hebeler-PRC85}.

\begin{figure}[H]
\begin{center}
\[
\begin{array}{ccc}
\includegraphics[width=2.15in]{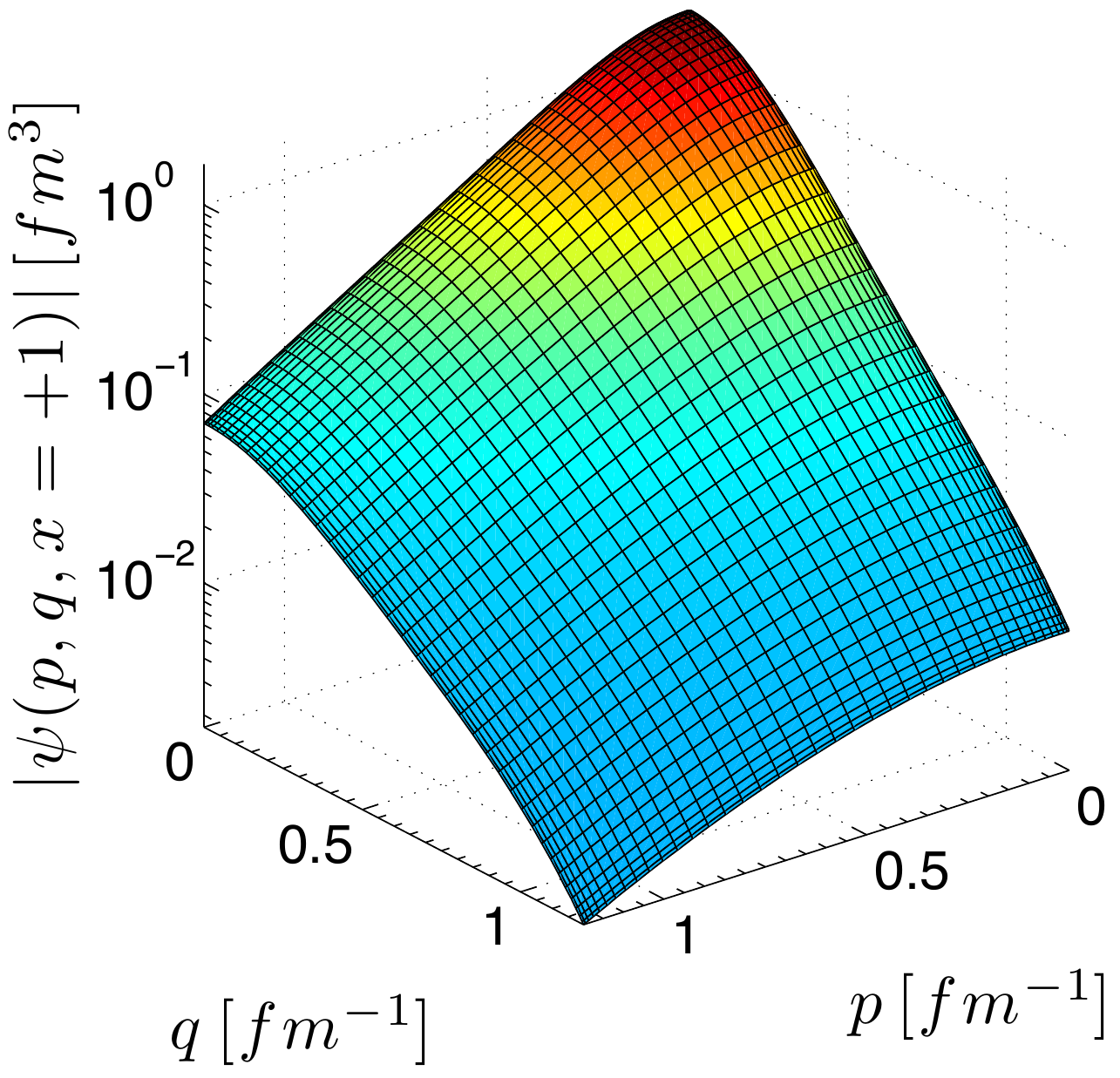}   &     
\hspace{0.5cm} 
\includegraphics[width=2.15in]{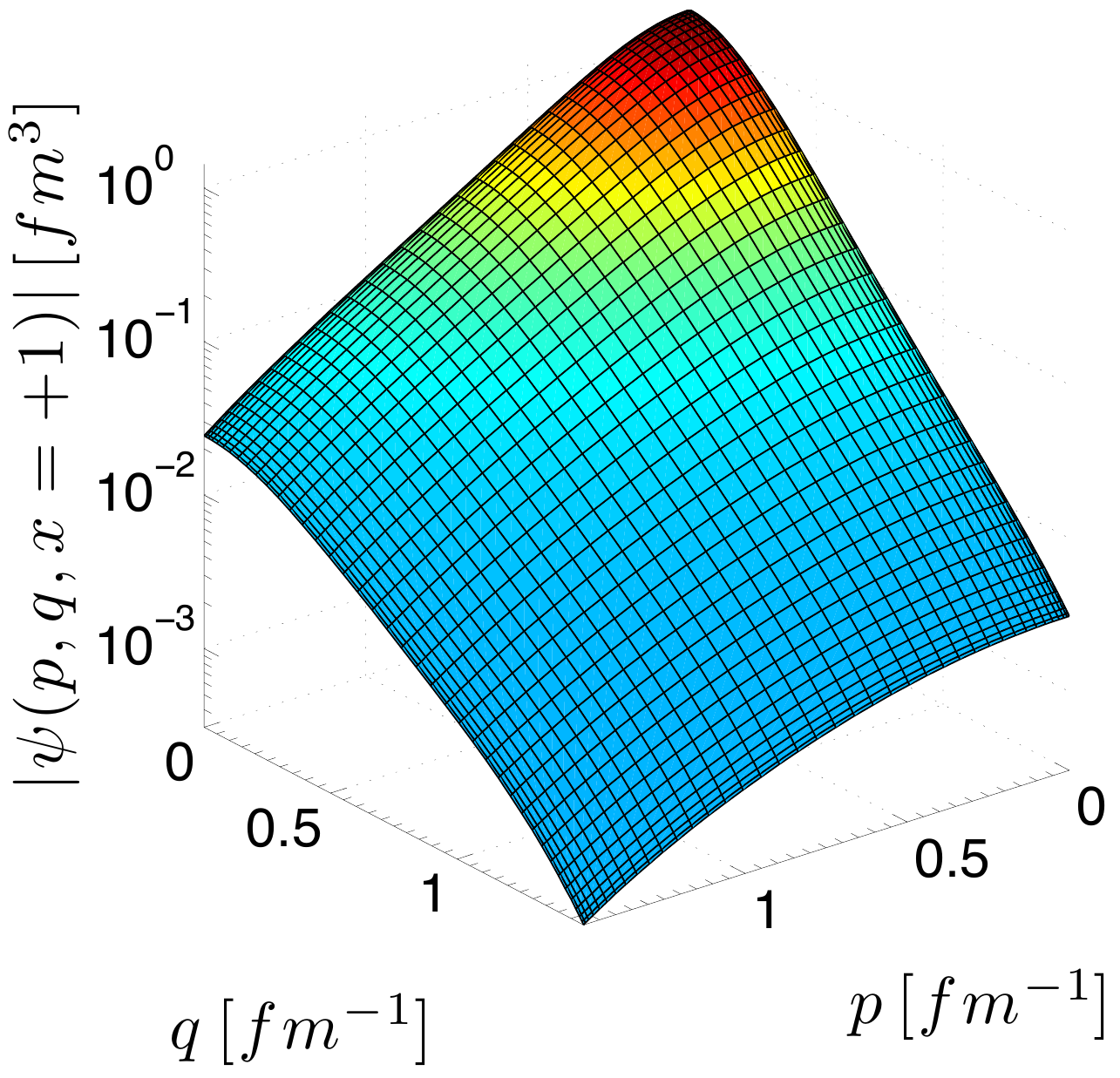} \\
\includegraphics[width=2.15in]{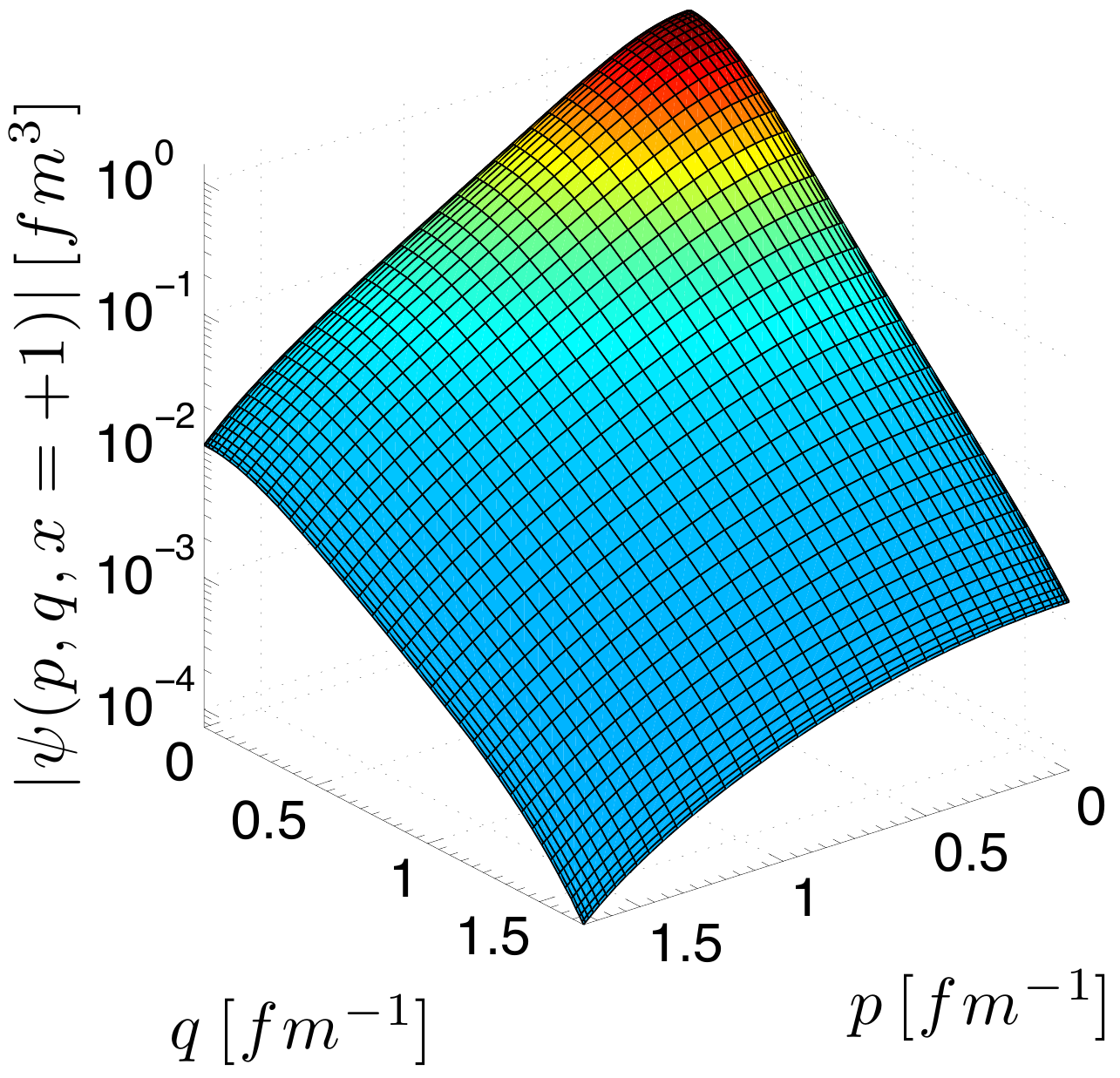}  &
\hspace{.5cm} 
\includegraphics[width=2.15in]{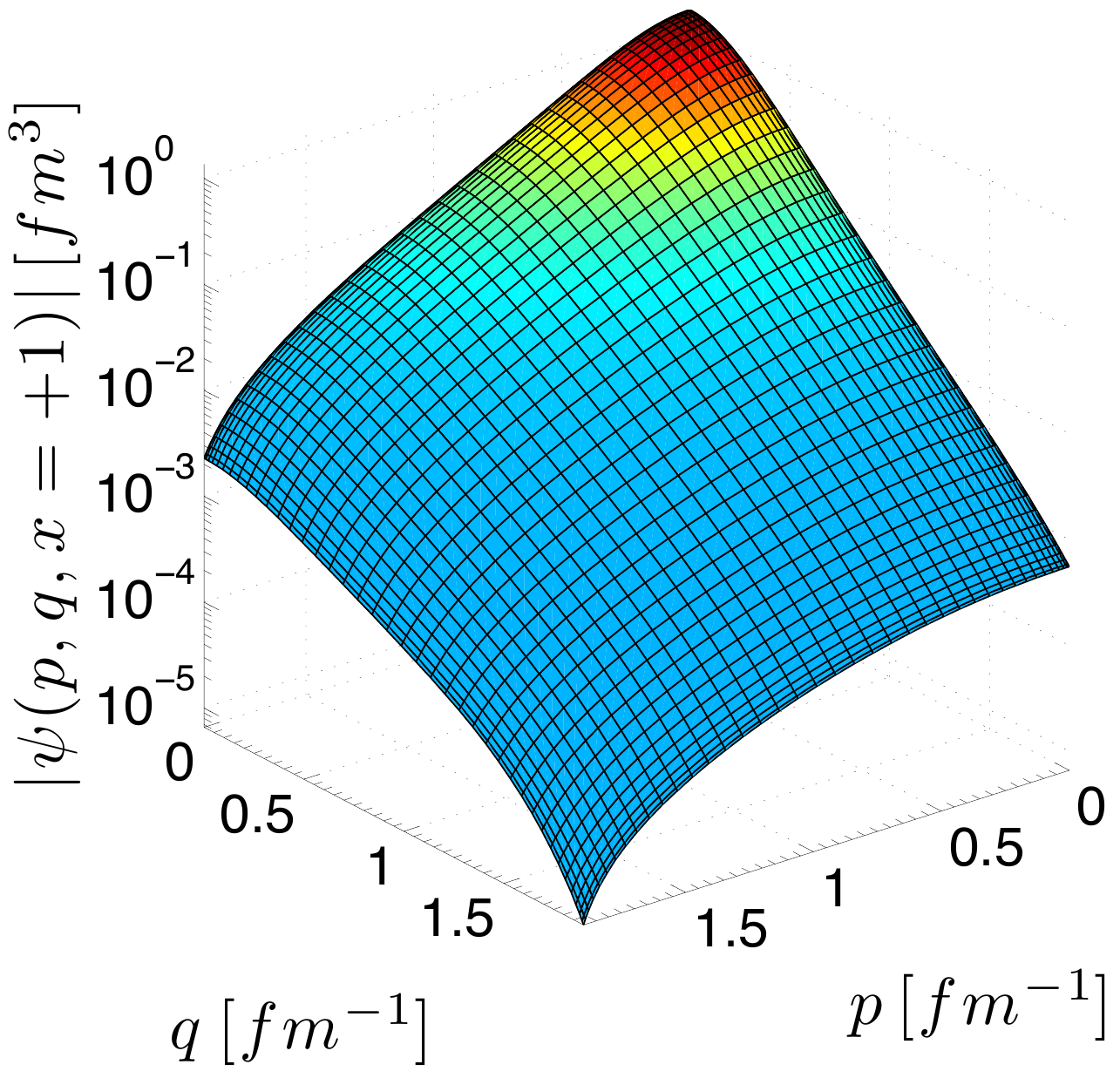}   \\   
\includegraphics[width=2.15in]{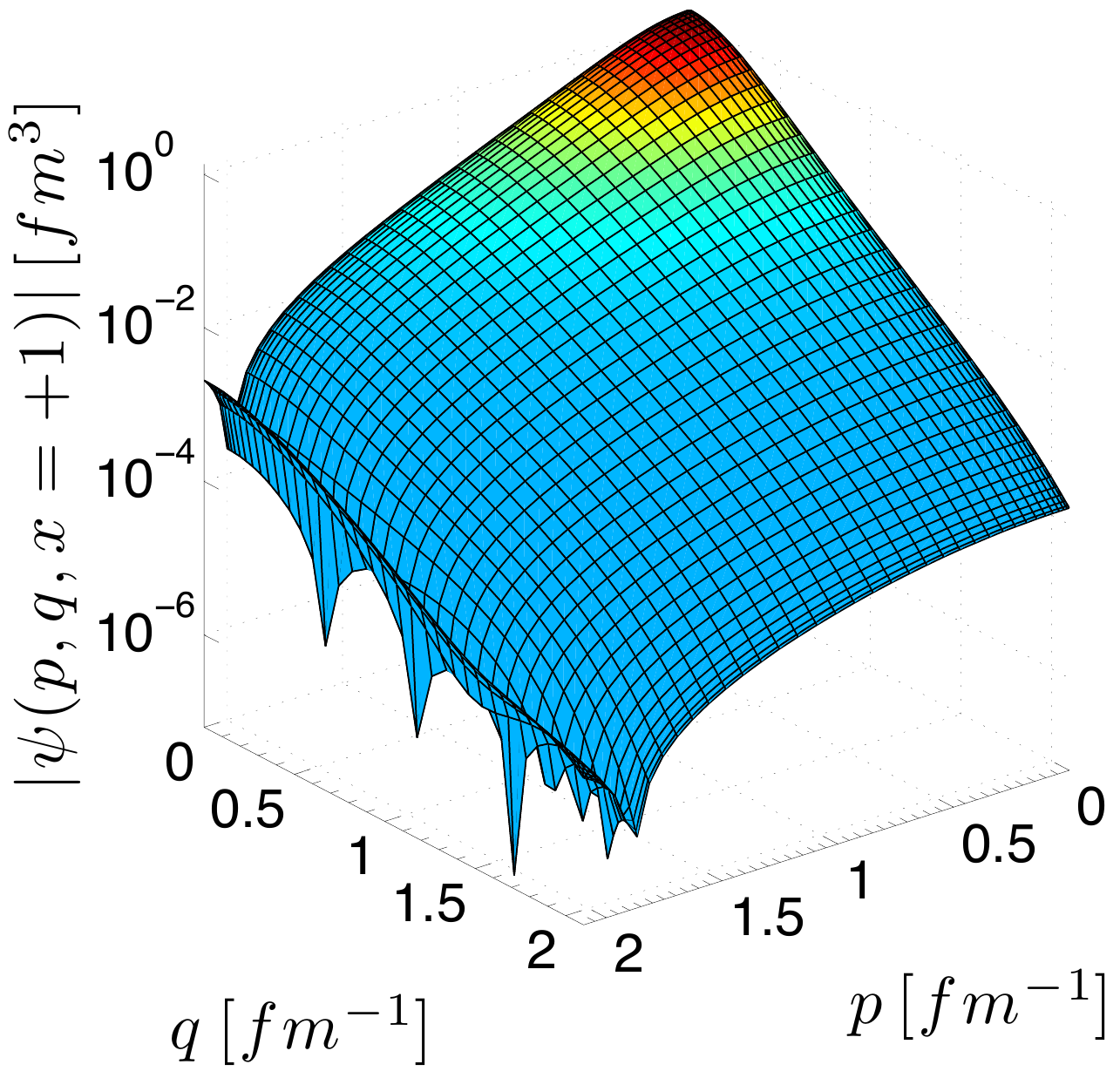} &
\hspace{.5cm}
\includegraphics[width=2.15in]{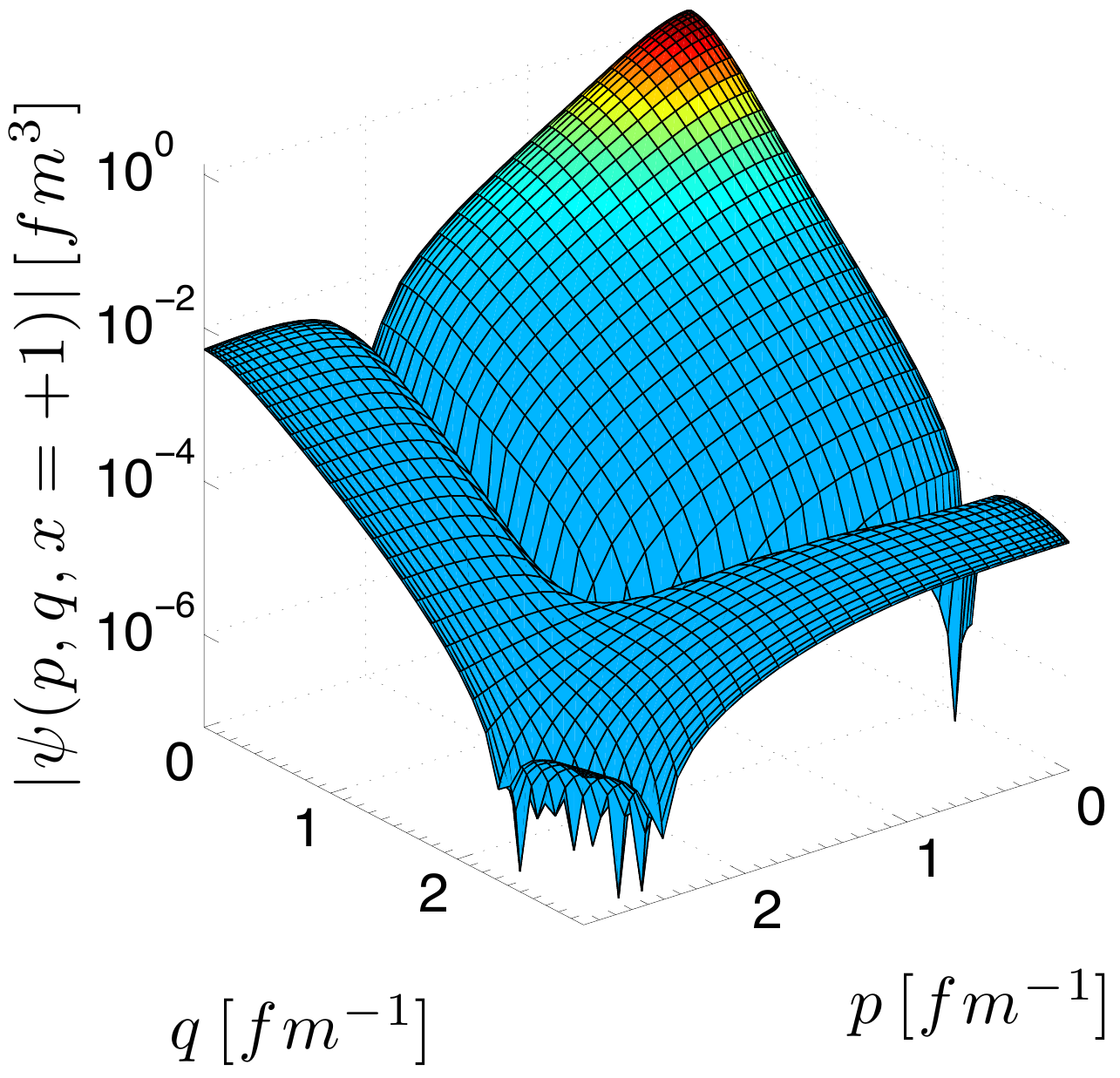}  \\
\includegraphics[width=2.15in]{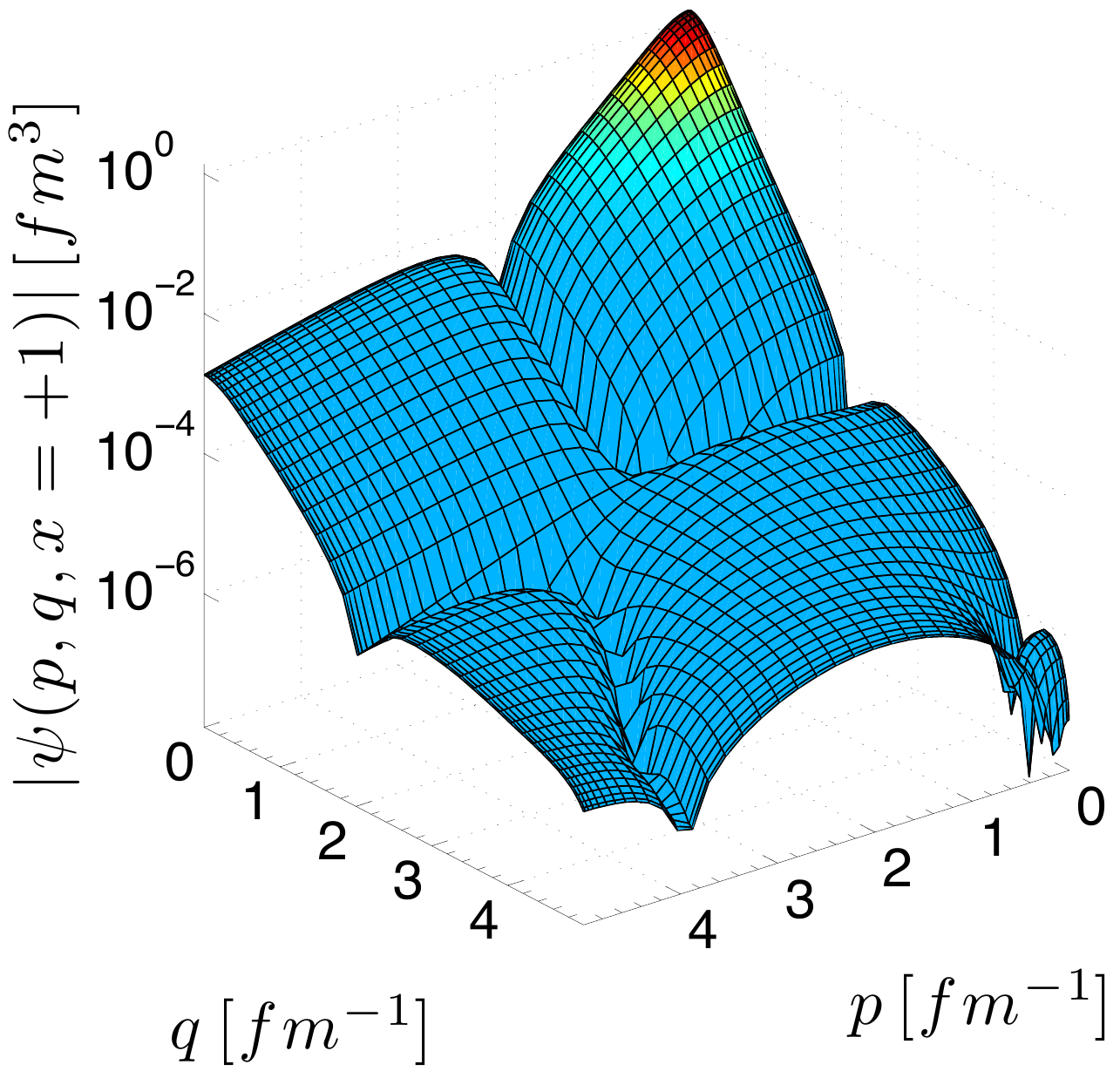}   &     
\hspace{.5cm} 
\includegraphics[width=2.15in]{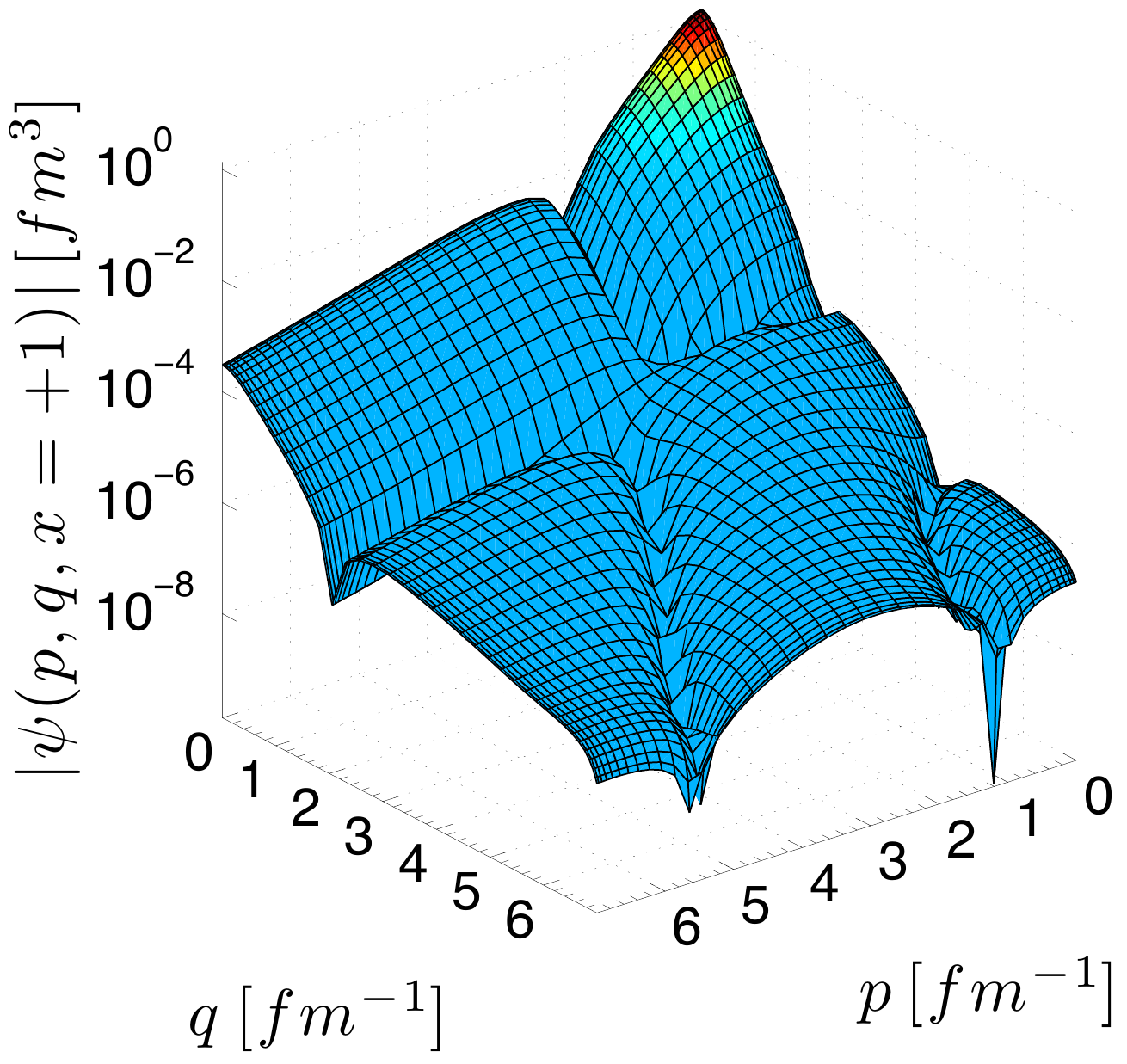} 
\end{array}
\]
\vspace{-1.cm} 
\end{center}
\caption{The magnitude of the Faddeev component $\psi(p,q,x)$ for $x=+1$ calculated from MT-V low momentum potential for $\Lambda=1.2-7.0\, fm^{-1}$.}
\label{fig:FC-3D-log}
\end{figure}

\begin{figure}[H]
\begin{center}
\[
\begin{array}{ccc}
\includegraphics[width=2.15in]{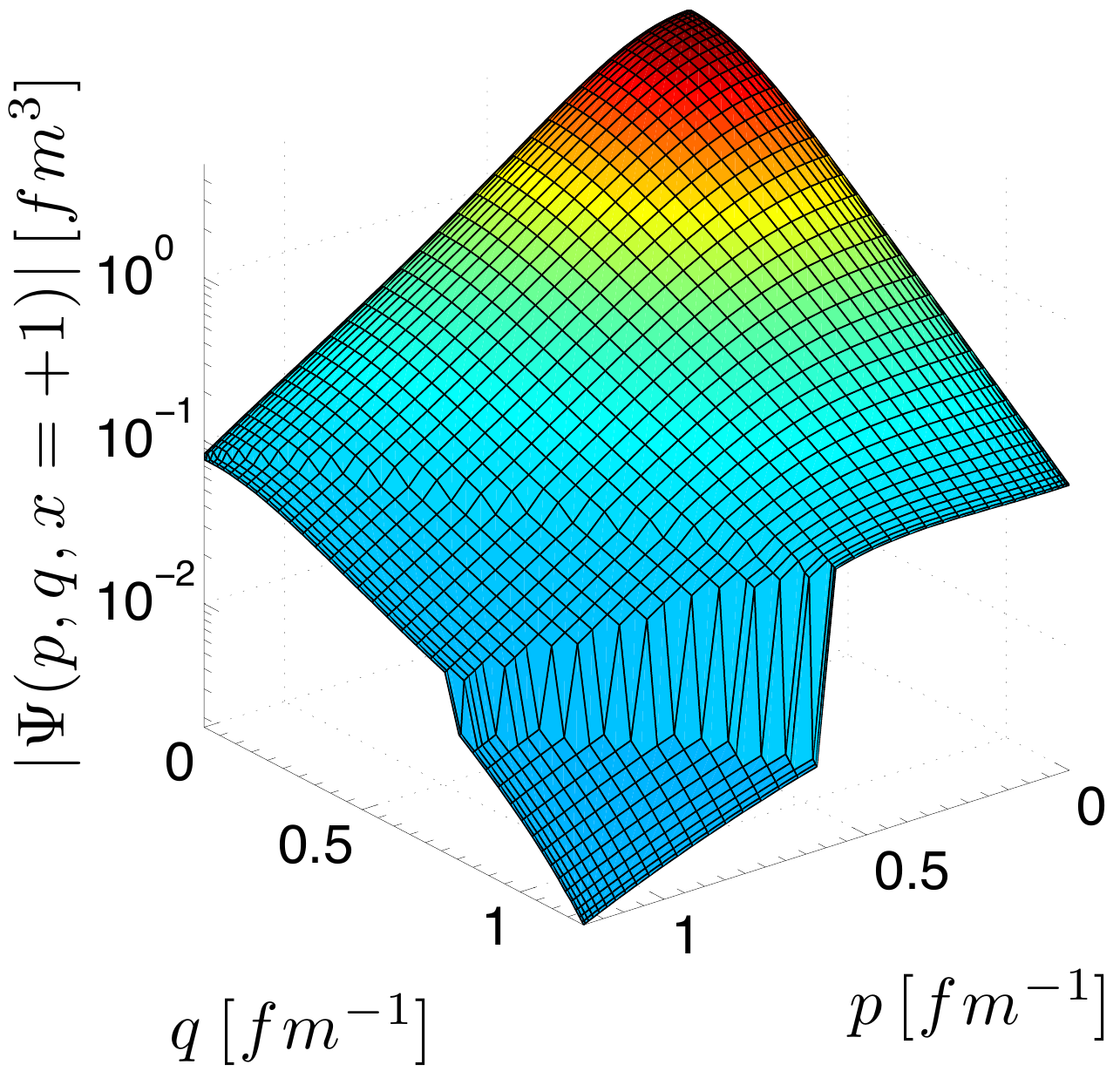}   &     
\hspace{0.5cm} 
\includegraphics[width=2.15in]{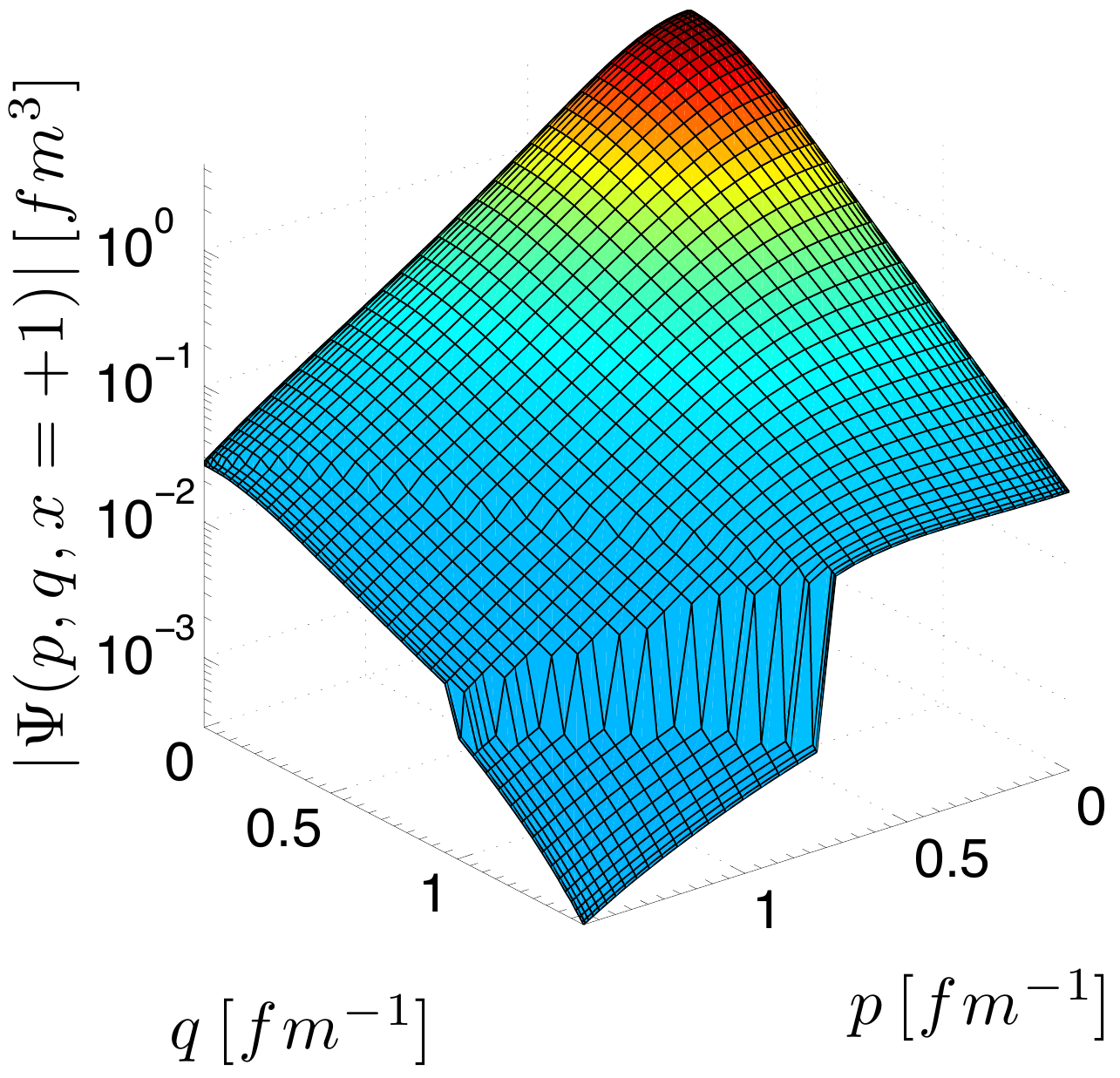} \\
\includegraphics[width=2.15in]{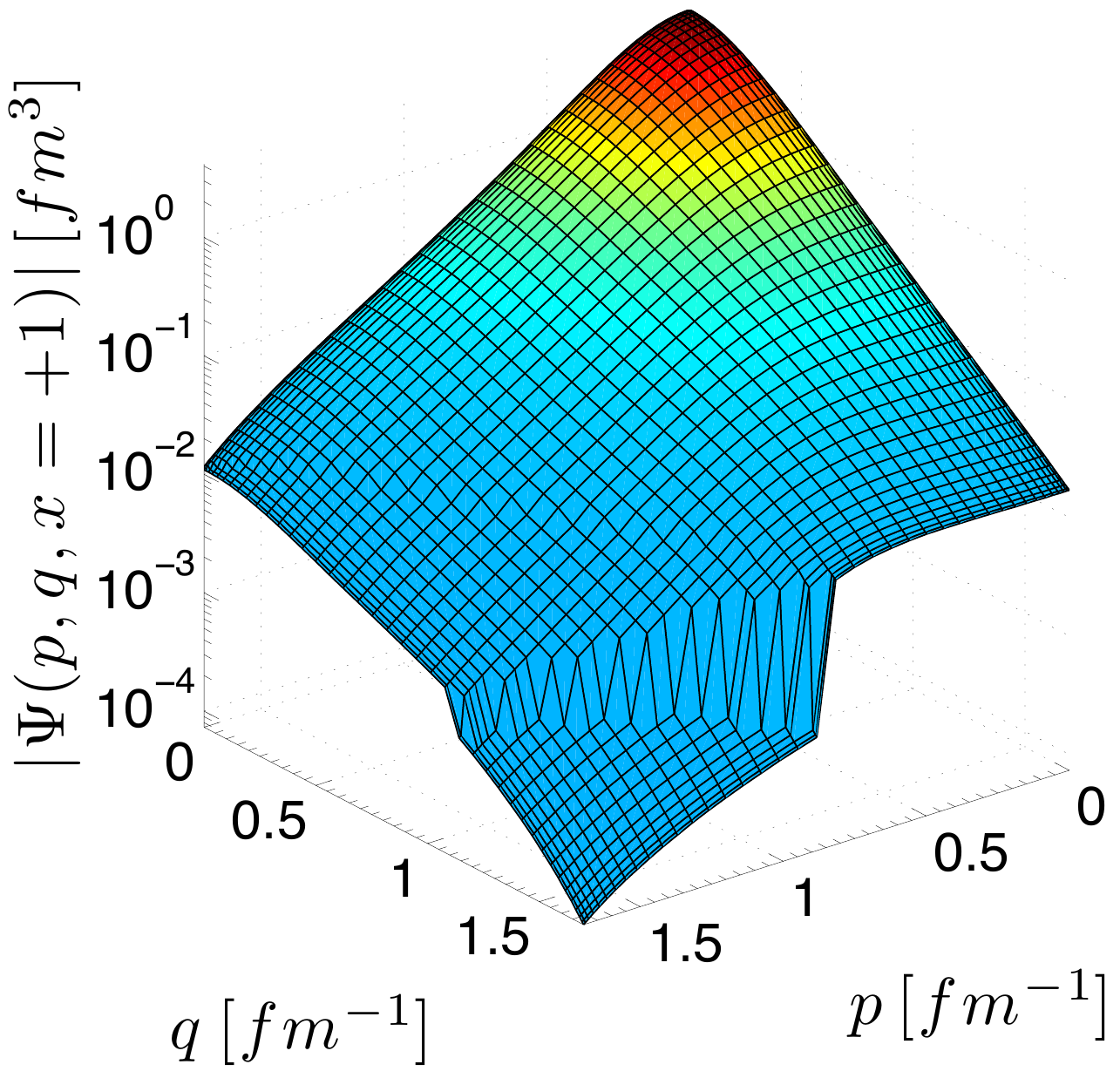}  &
\hspace{.5cm} 
\includegraphics[width=2.15in]{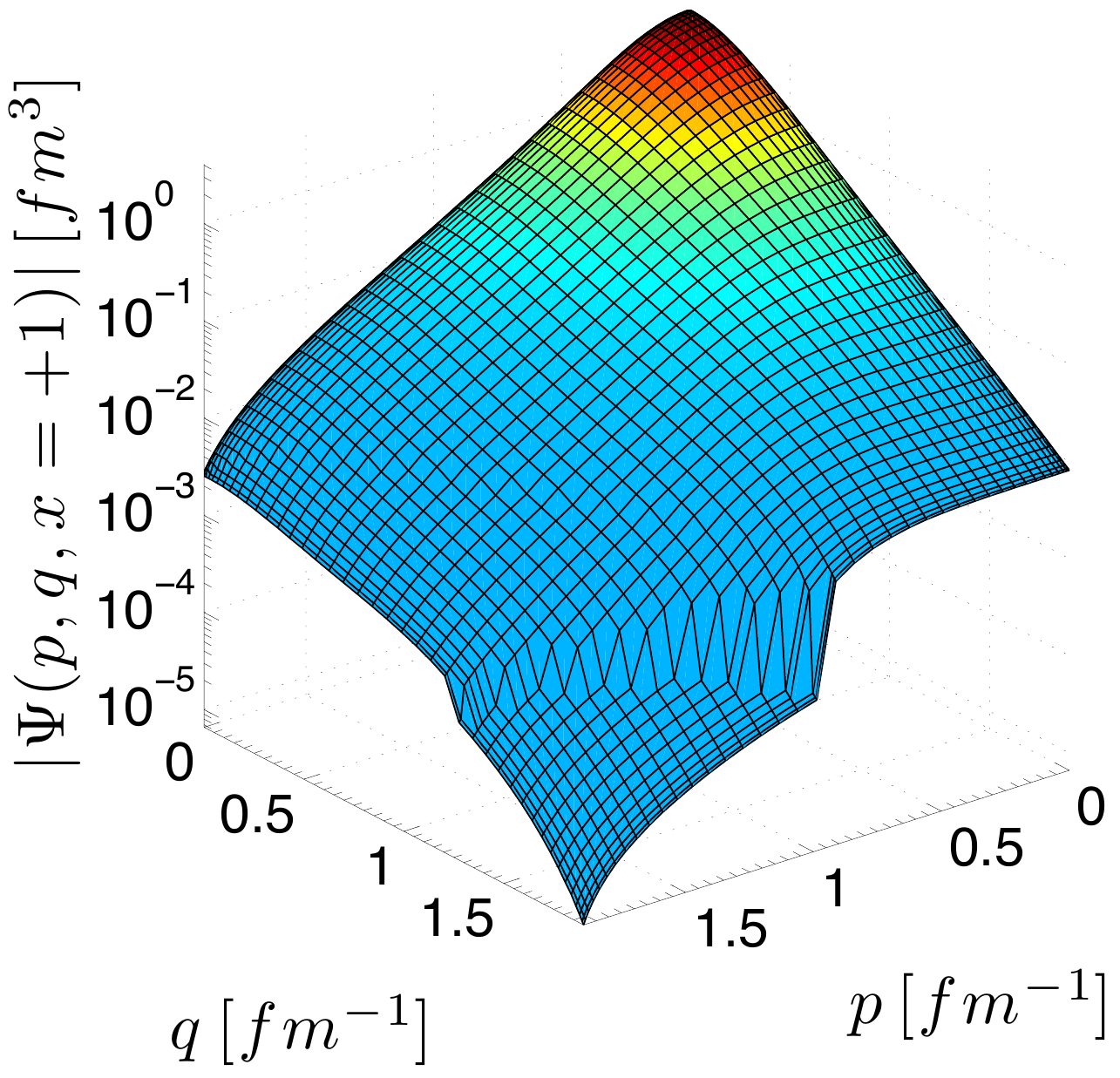}   \\   
\includegraphics[width=2.15in]{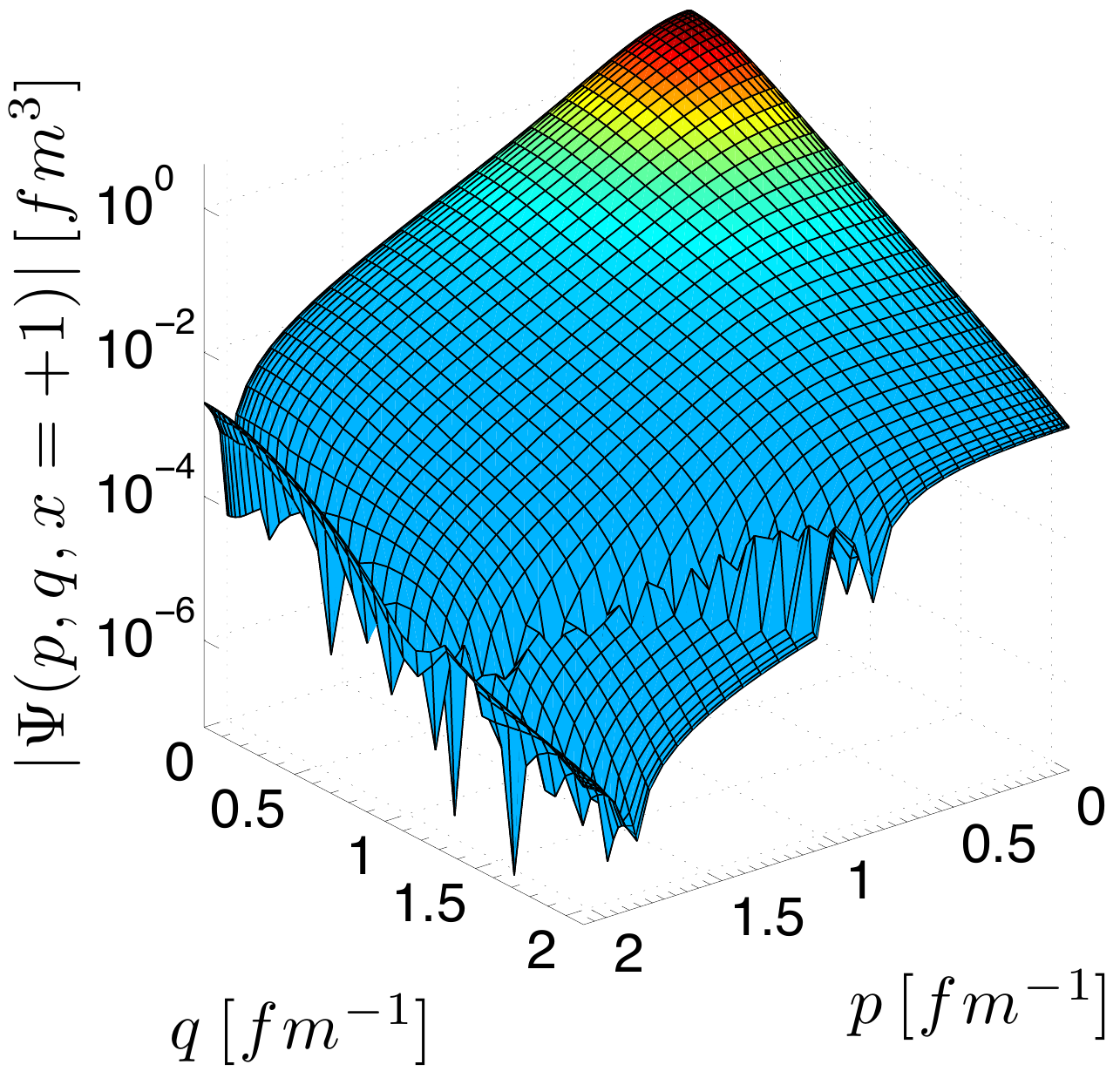} &
\hspace{.5cm}
\includegraphics[width=2.15in]{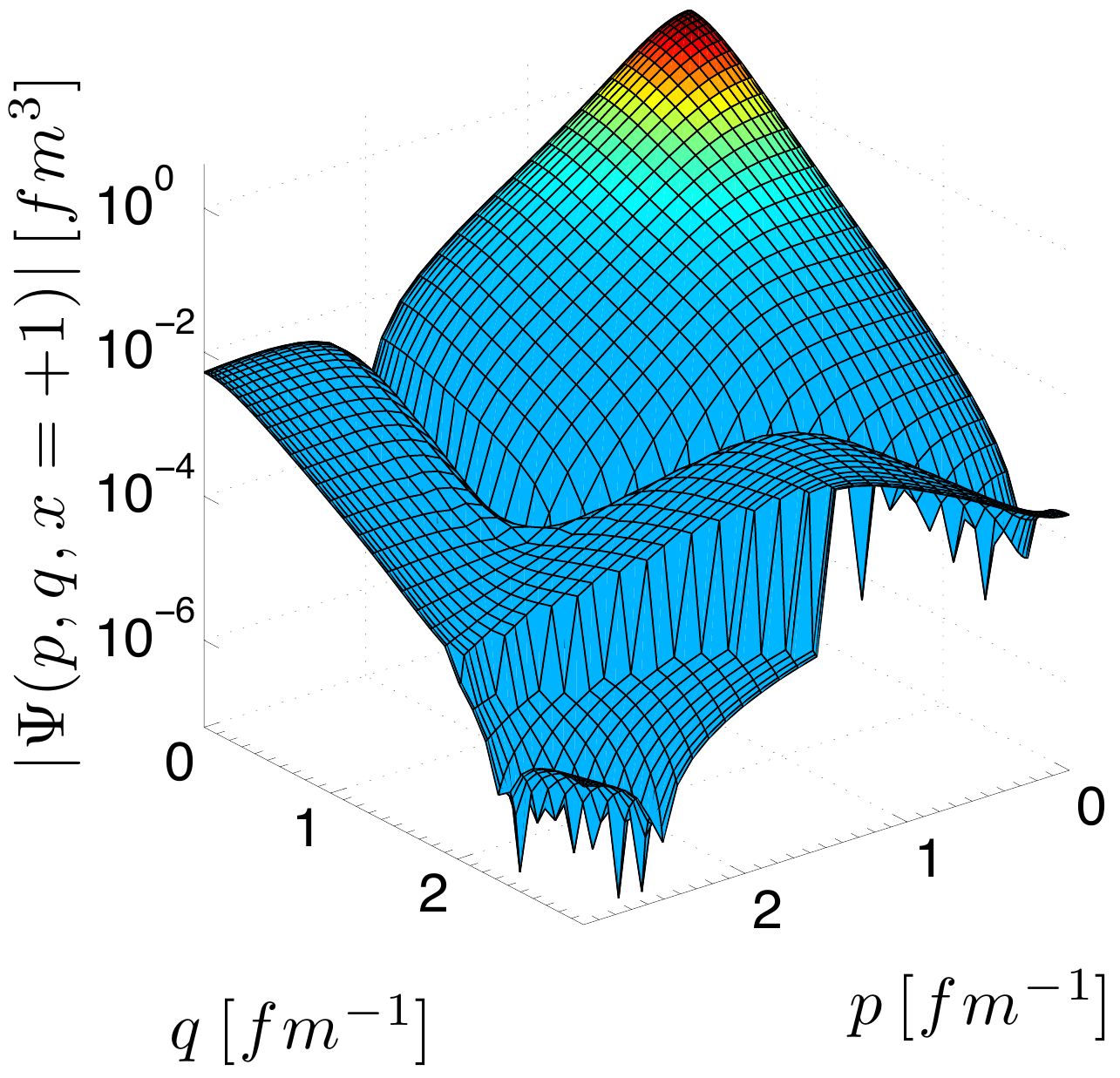}  \\
\includegraphics[width=2.15in]{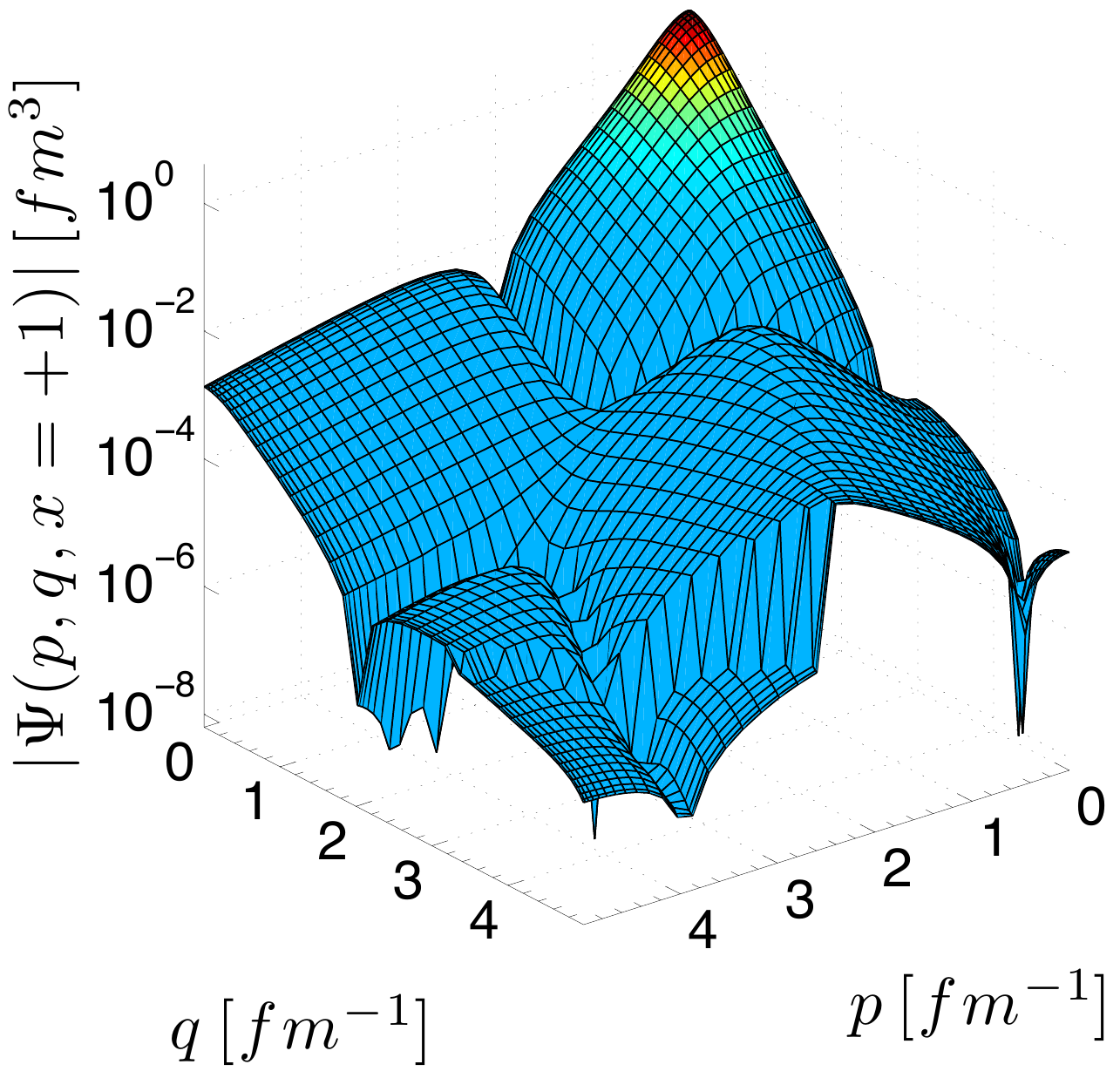}   &     
\hspace{.5cm} 
\includegraphics[width=2.15in]{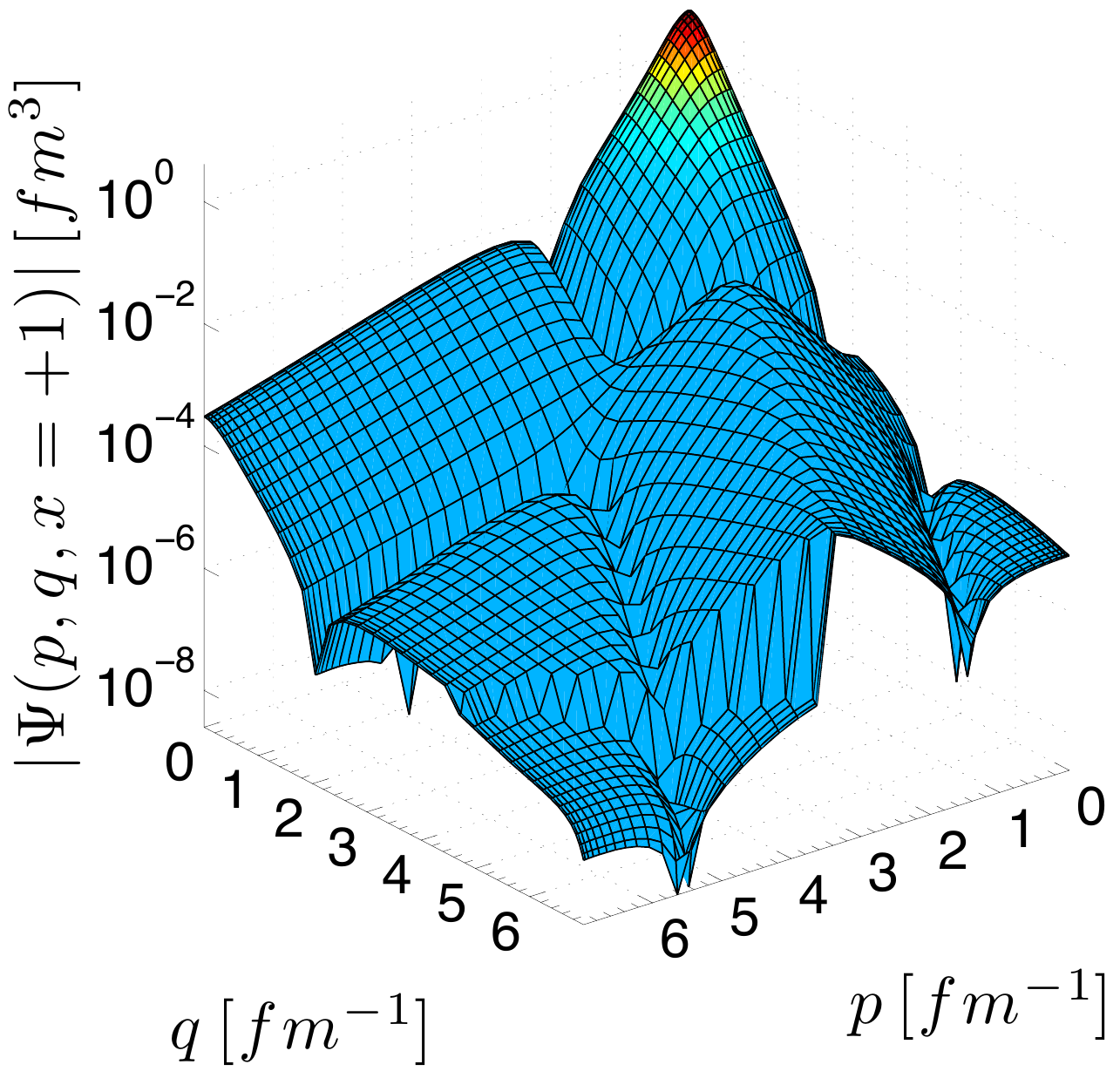} 
\end{array}
\]
\end{center}
\caption{The magnitude of the 3B bound state wave function $\Psi(p,q,x)$ for $x=+1$ calculated from MT-V low momentum potential for $\Lambda=1.2-7.0\, fm^{-1}$.}
\label{fig:WF-3D-log}
\end{figure}

In Figs. \ref{fig:FC-3D-log} and \ref{fig:WF-3D-log}, we have shown the Faddeev components $\psi(p,q,x)$ and total wave functions $\Psi(p,q,x)$ for cutoffs $\Lambda=1.2, 1.5, 1.7, 1.9, 2.1, 3.0, 5.0, 7.0\, fm^{-1}$ for fixed angle $x=+1$ on a logarithmic scale on $z$-axis. 
These figures clearly show that when the cutoff $\Lambda$ is decreased to small values, the effect of hard core interaction disappears. This is a consequence of integrating out the short-distance physics, which presents a repulsive core, in construction of low momentum interaction.

\subsection{Momentum Probability Densities}\label{}

\begin{figure}[H]
\begin{center}
\[
\begin{array}{ccc}
\includegraphics[width=2.5in]{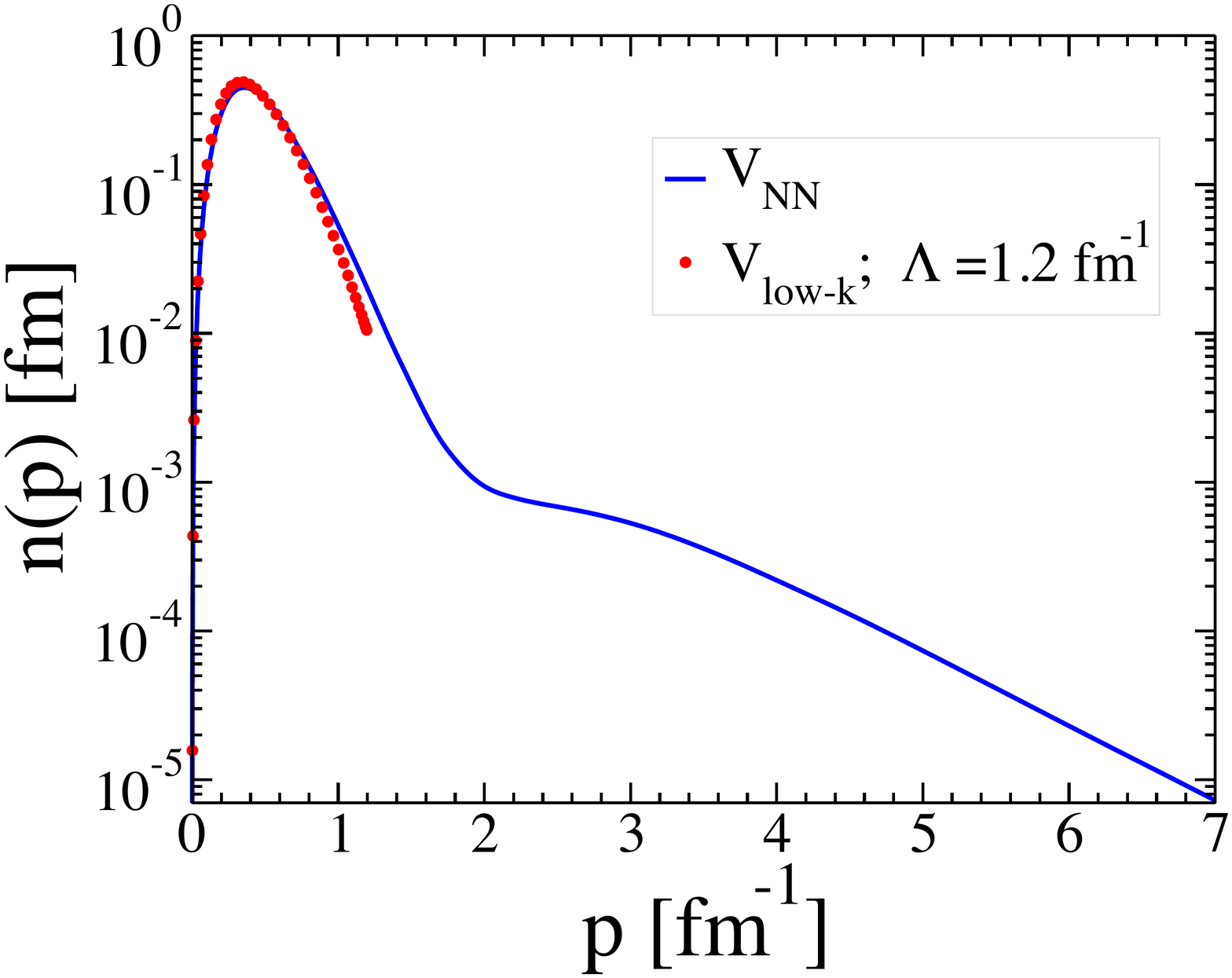}   &     
\hspace{.cm} 
\includegraphics[width=2.5in]{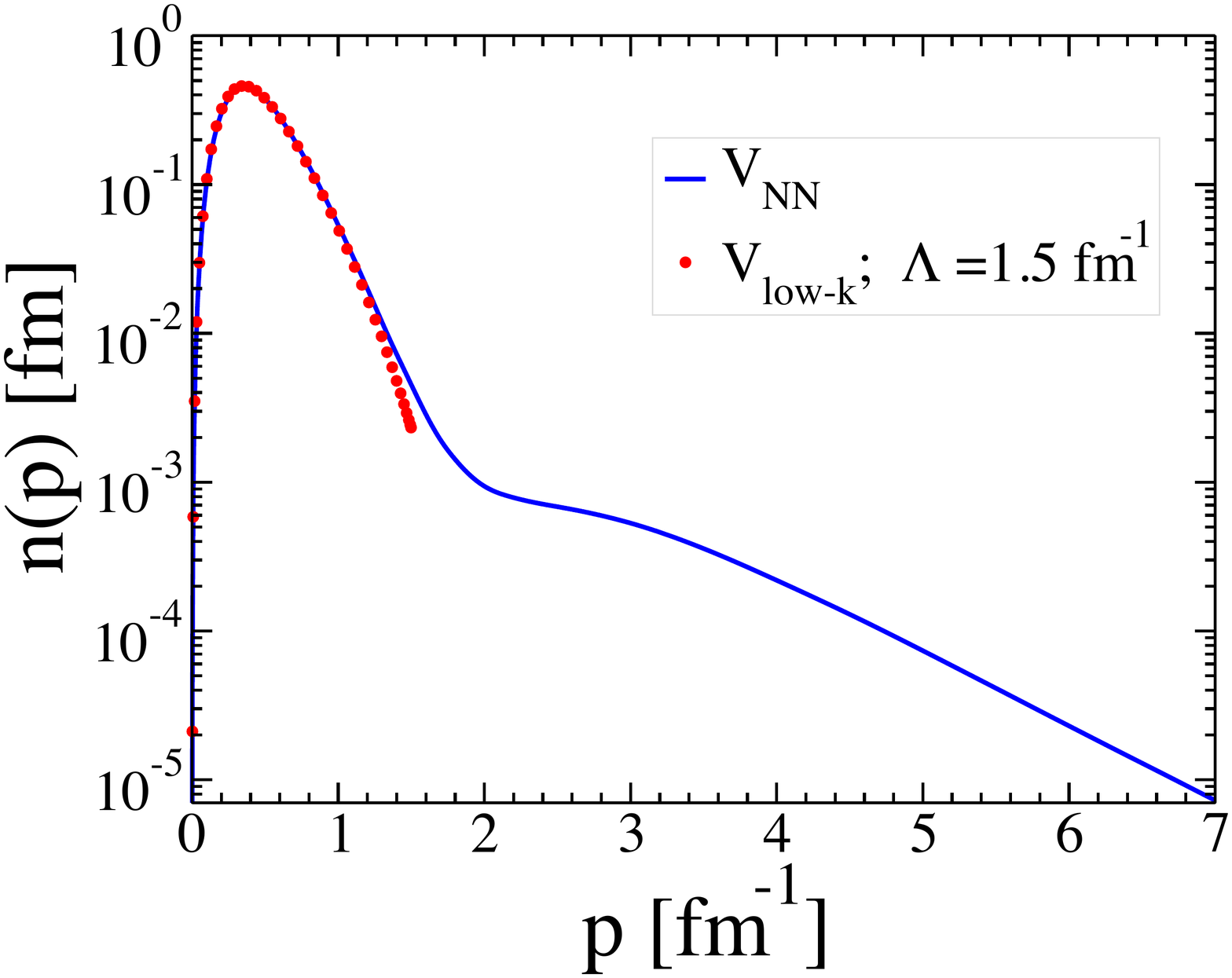} \\
\includegraphics[width=2.5in]{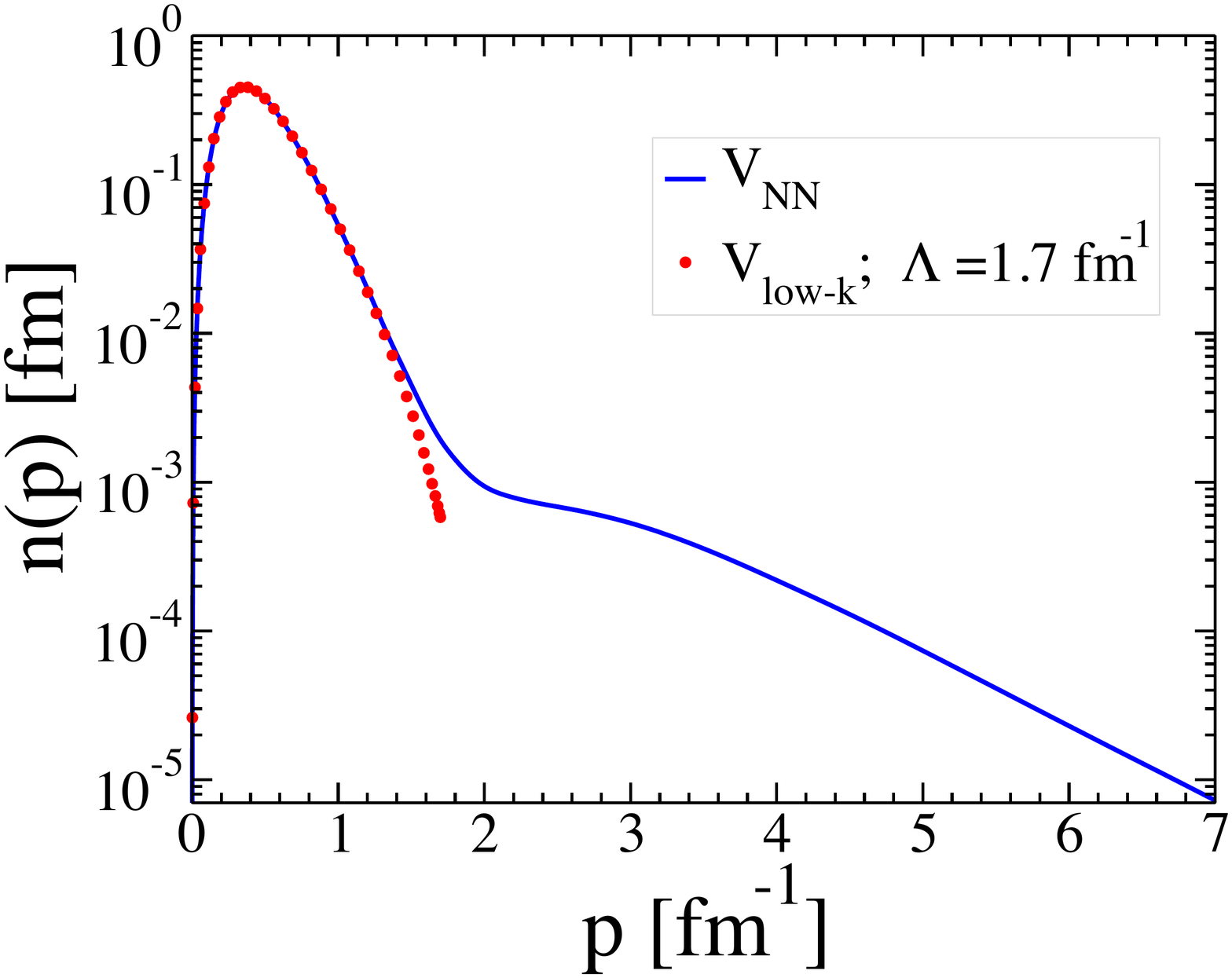}  &
\hspace{.cm}
\includegraphics[width=2.5in]{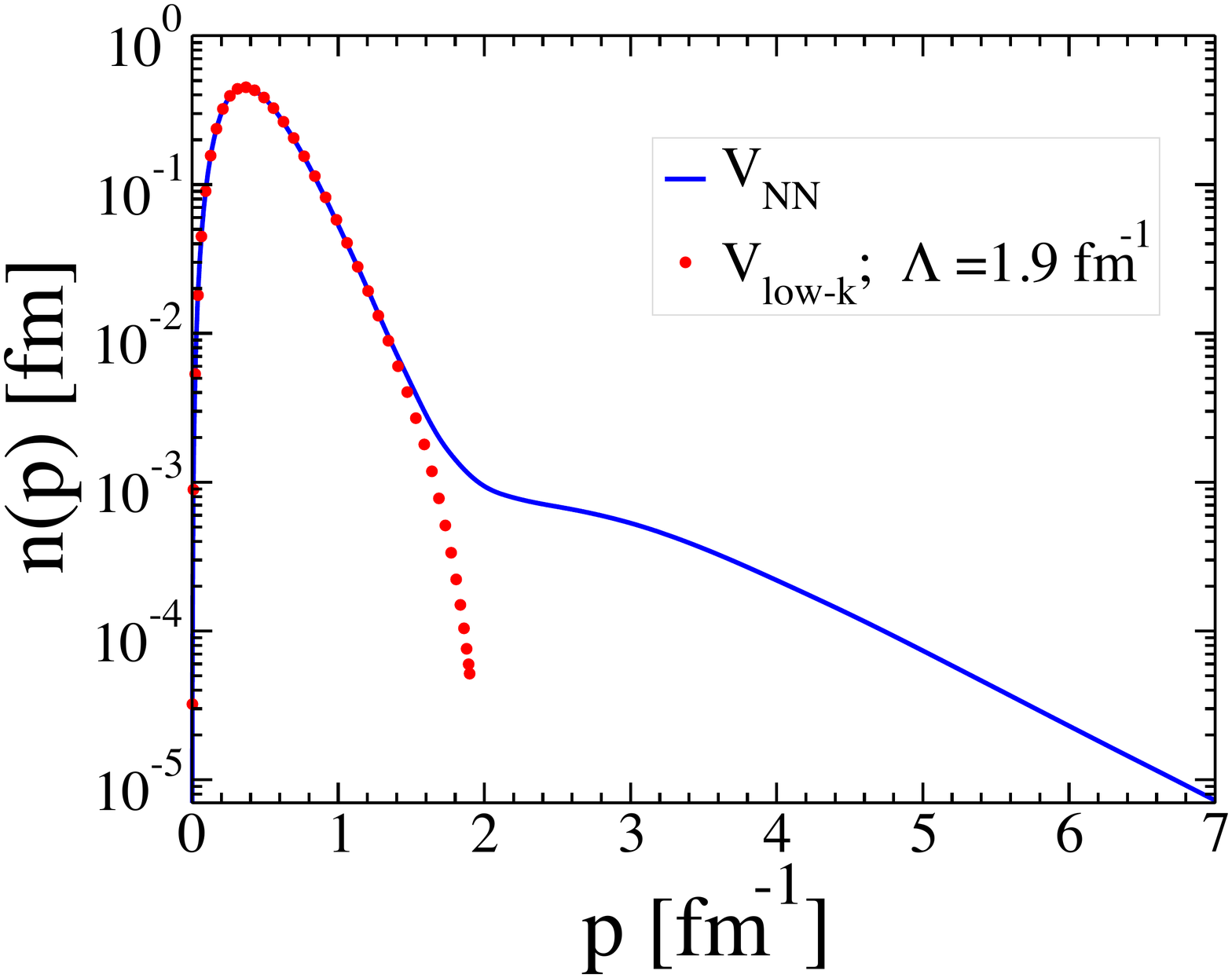}   \\   
\includegraphics[width=2.5in]{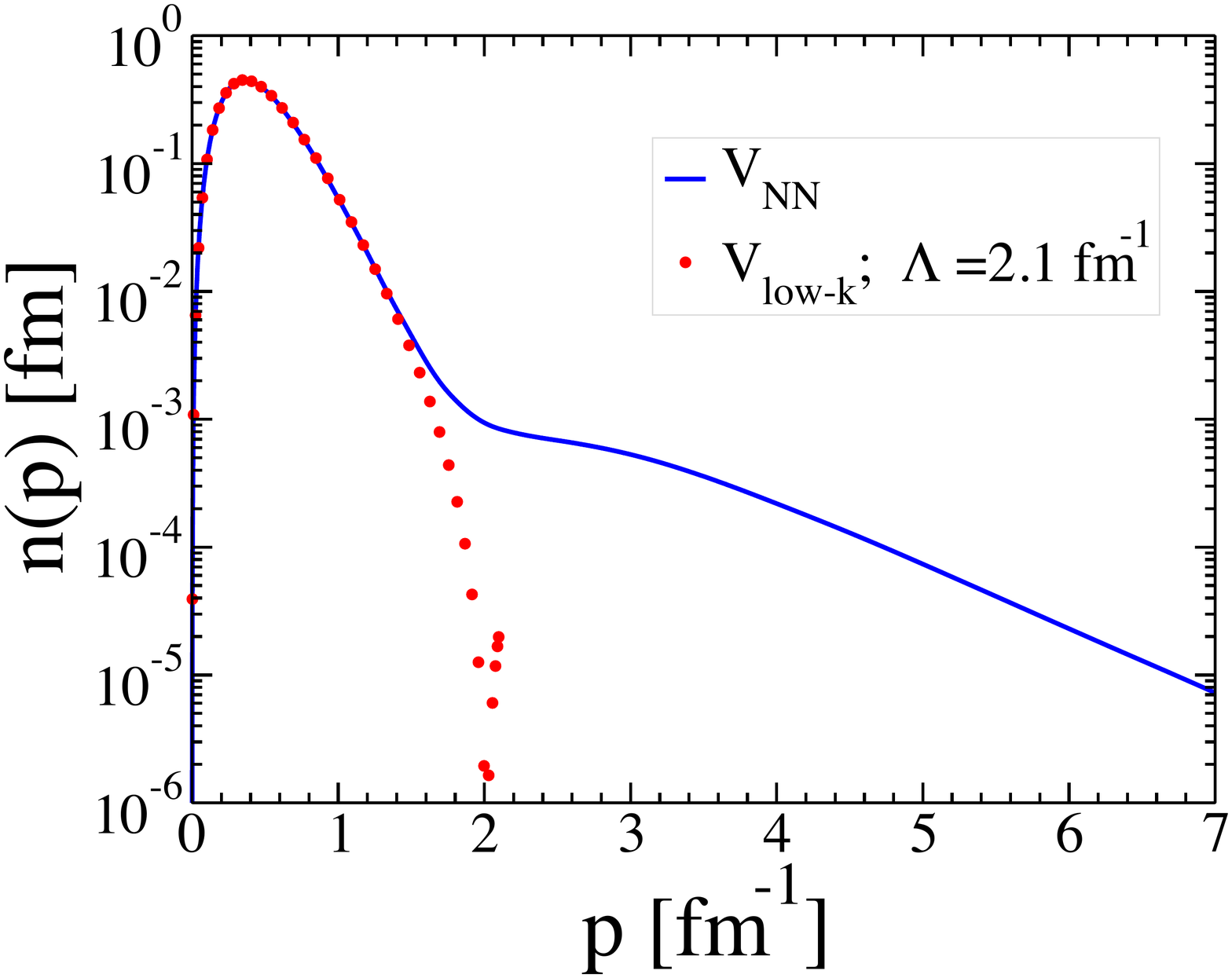} &
\hspace{.cm}
\includegraphics[width=2.5in]{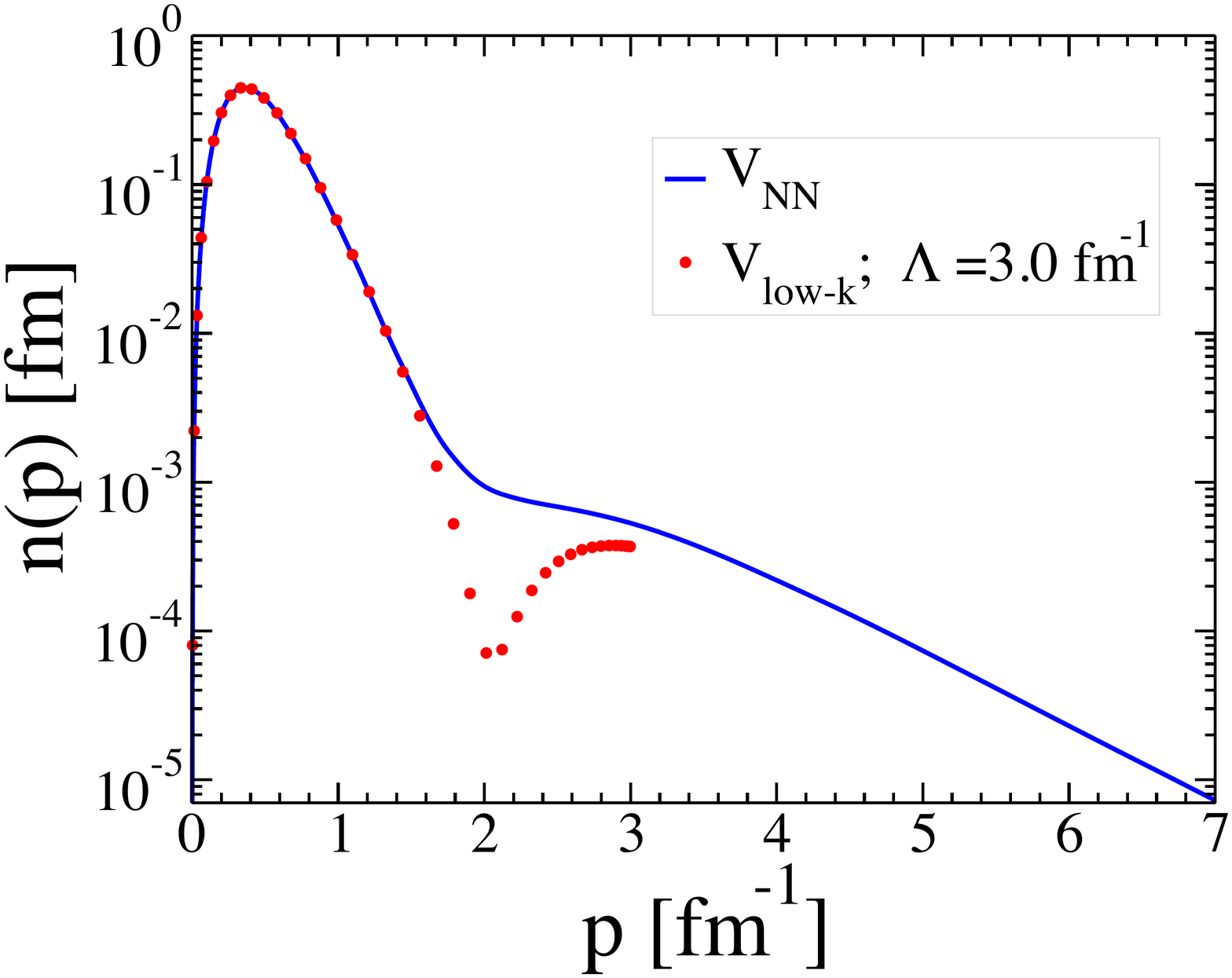}  \\
\includegraphics[width=2.5in]{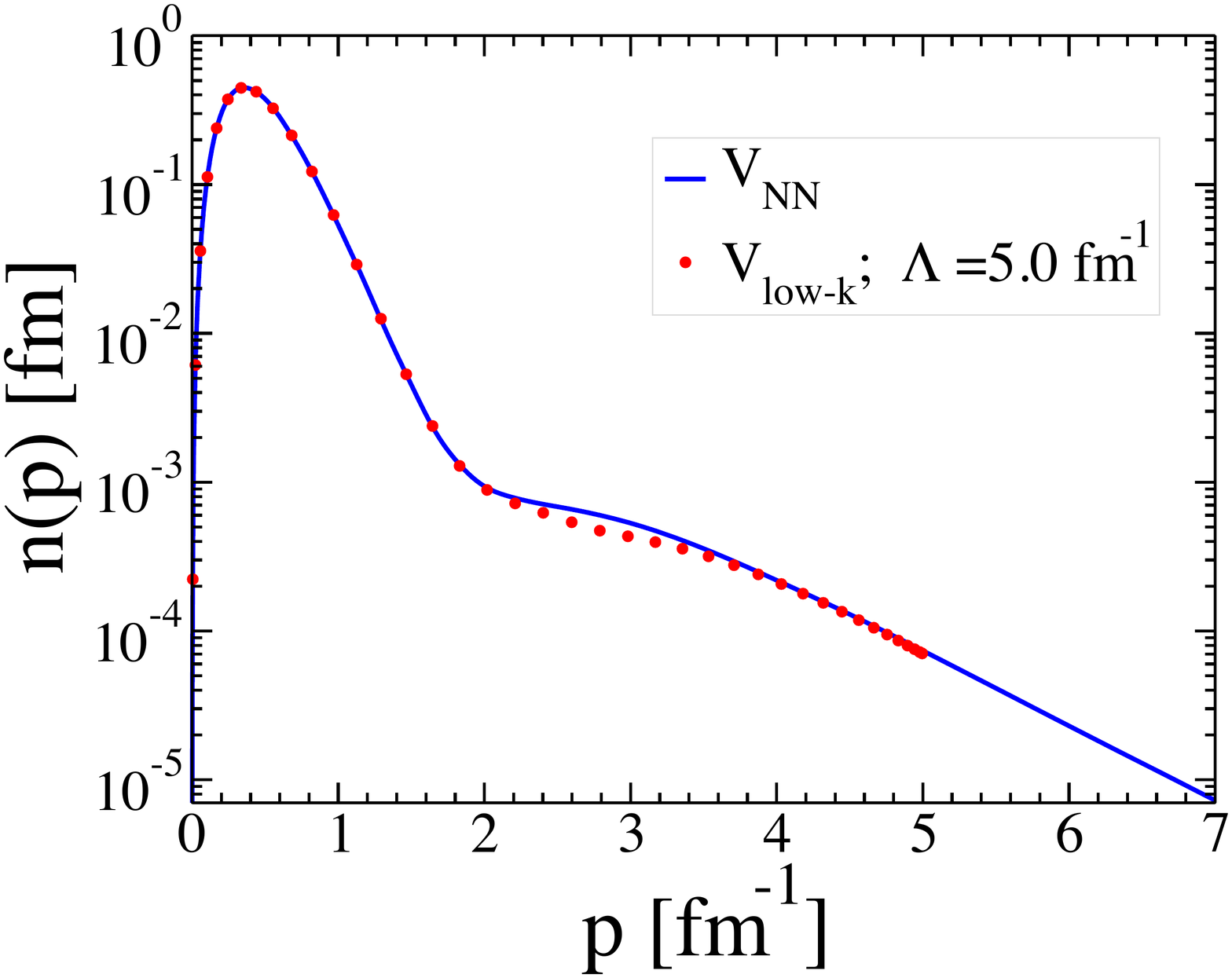}   &     
\hspace{.cm} 
\includegraphics[width=2.5in]{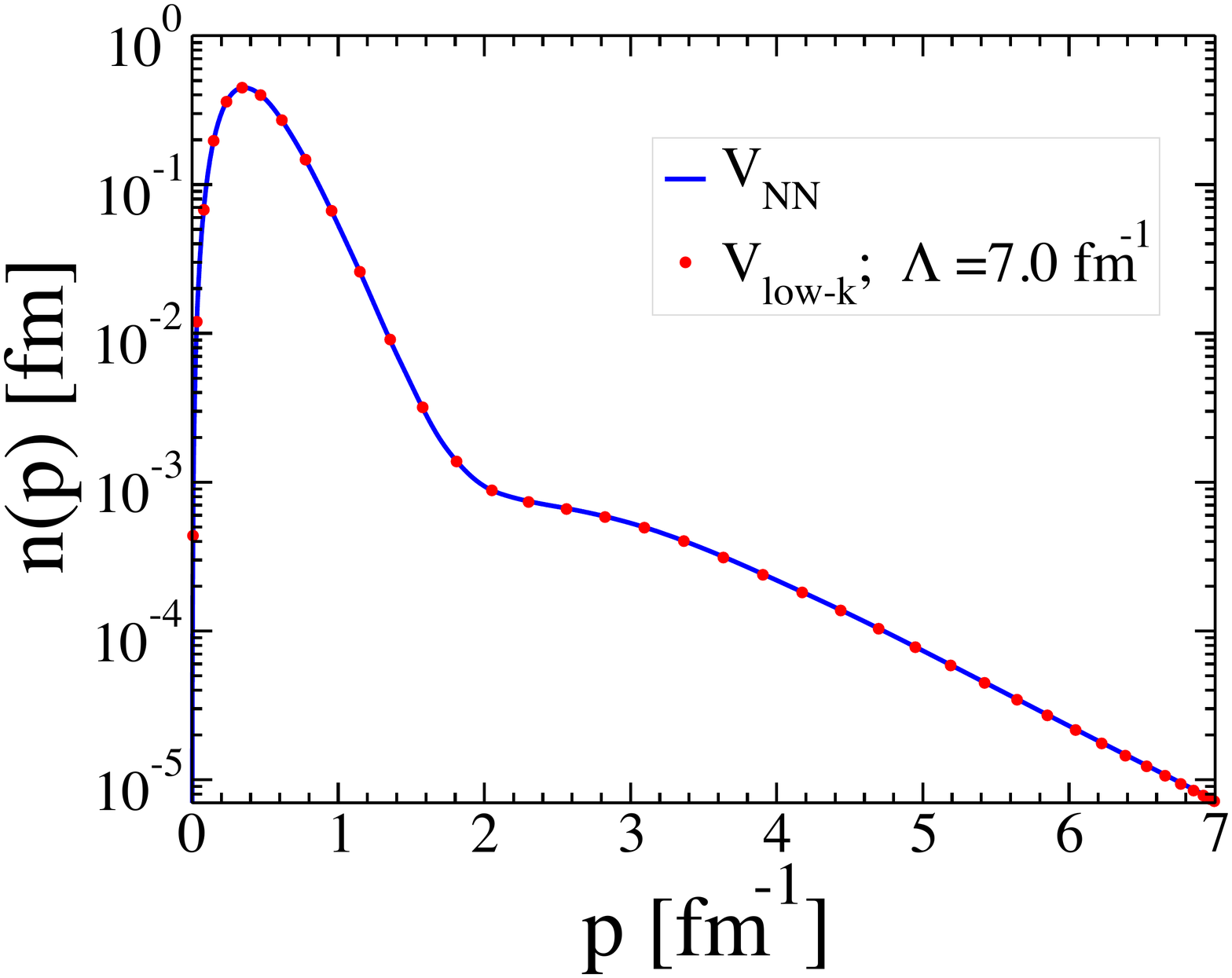} 
\end{array}
\]
\end{center}
\caption{ The momentum distribution function $n(p)$ calculated from MT-V bare and low momentum potential for $\Lambda=1.2-7.0\, fm^{-1}$}
\label{fig:np}
\end{figure}

\begin{figure}[H]
\begin{center}
\[
\begin{array}{ccc}
\includegraphics[width=2.5in]{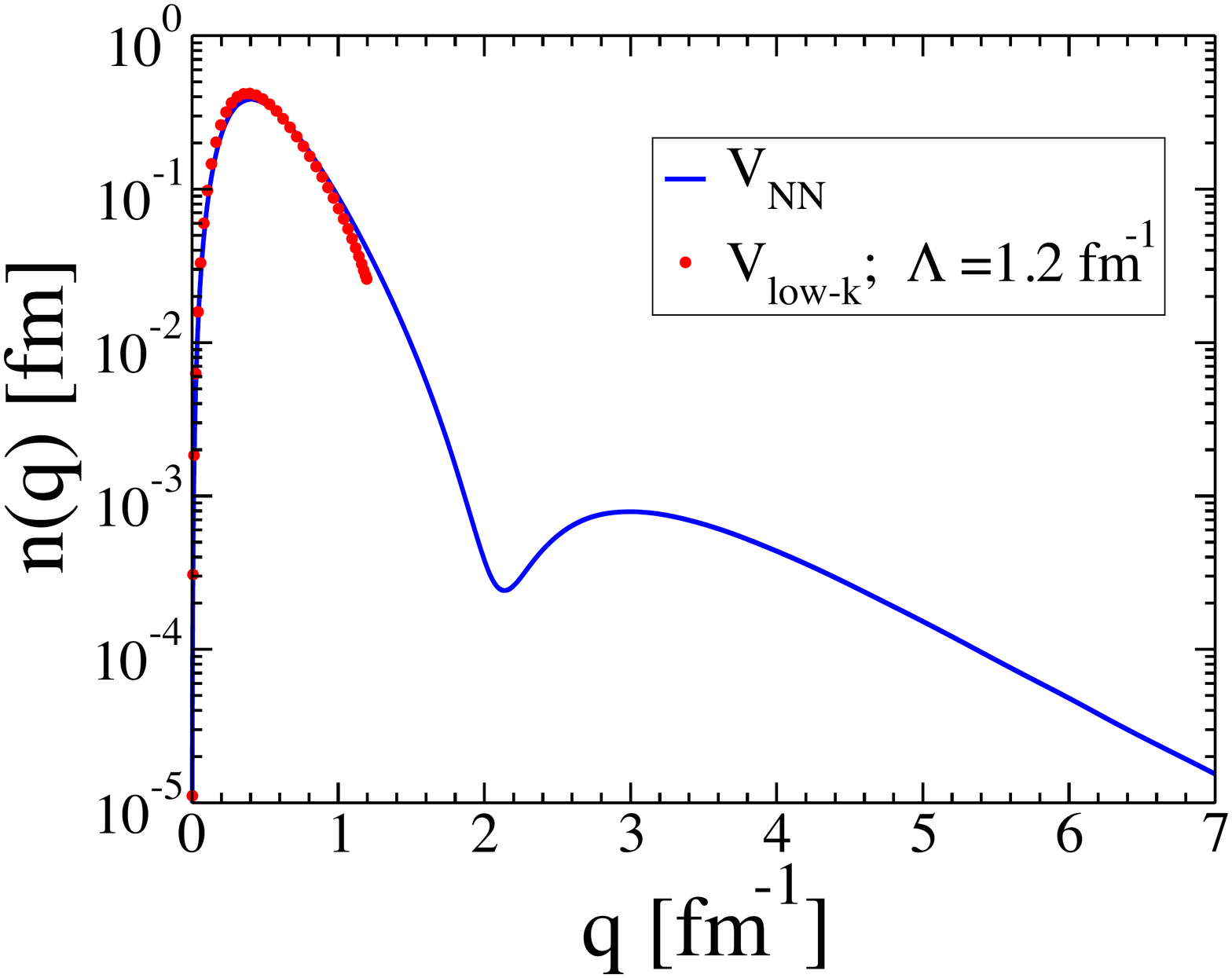}   &     
\hspace{.cm} 
\includegraphics[width=2.5in]{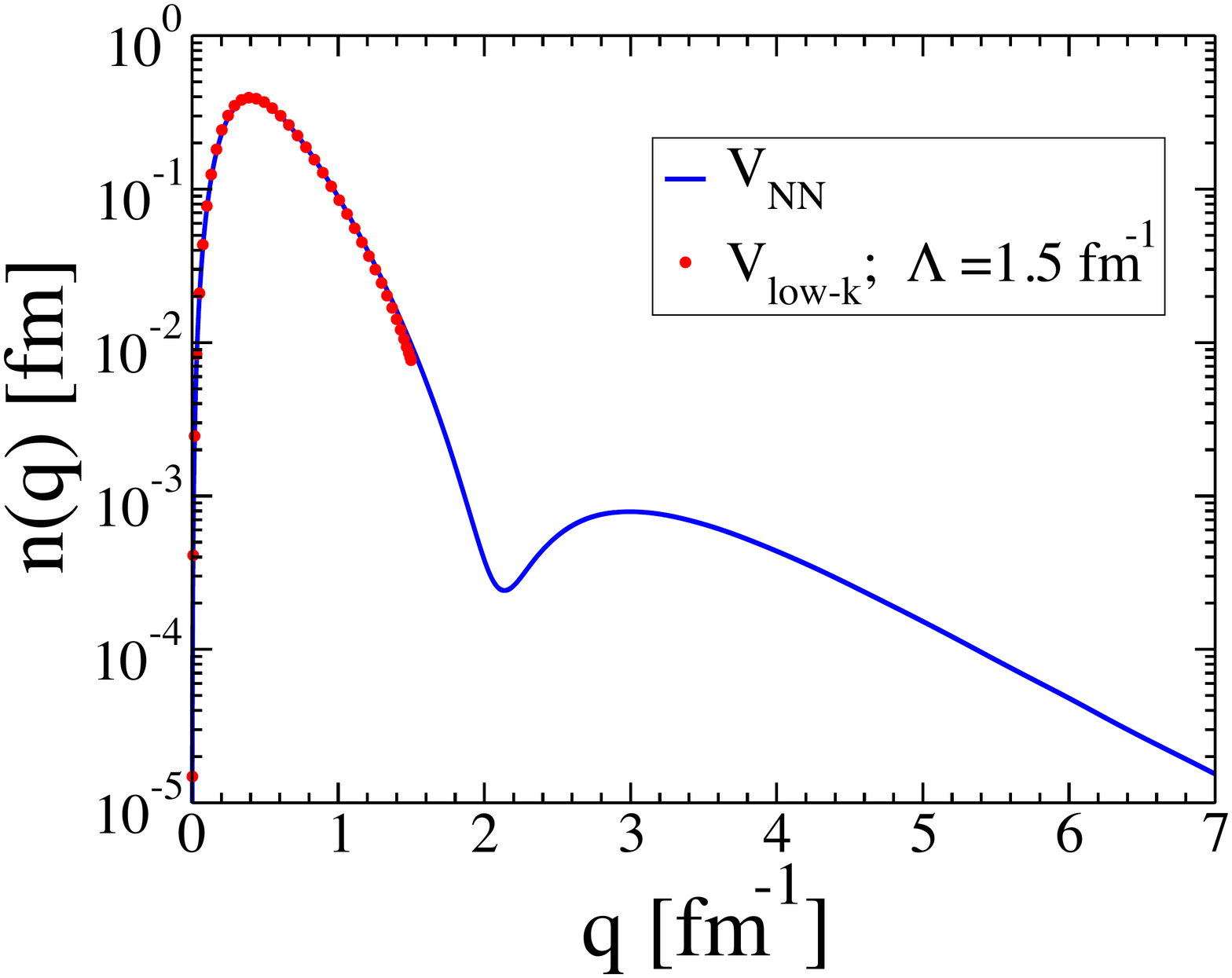} \\
\includegraphics[width=2.5in]{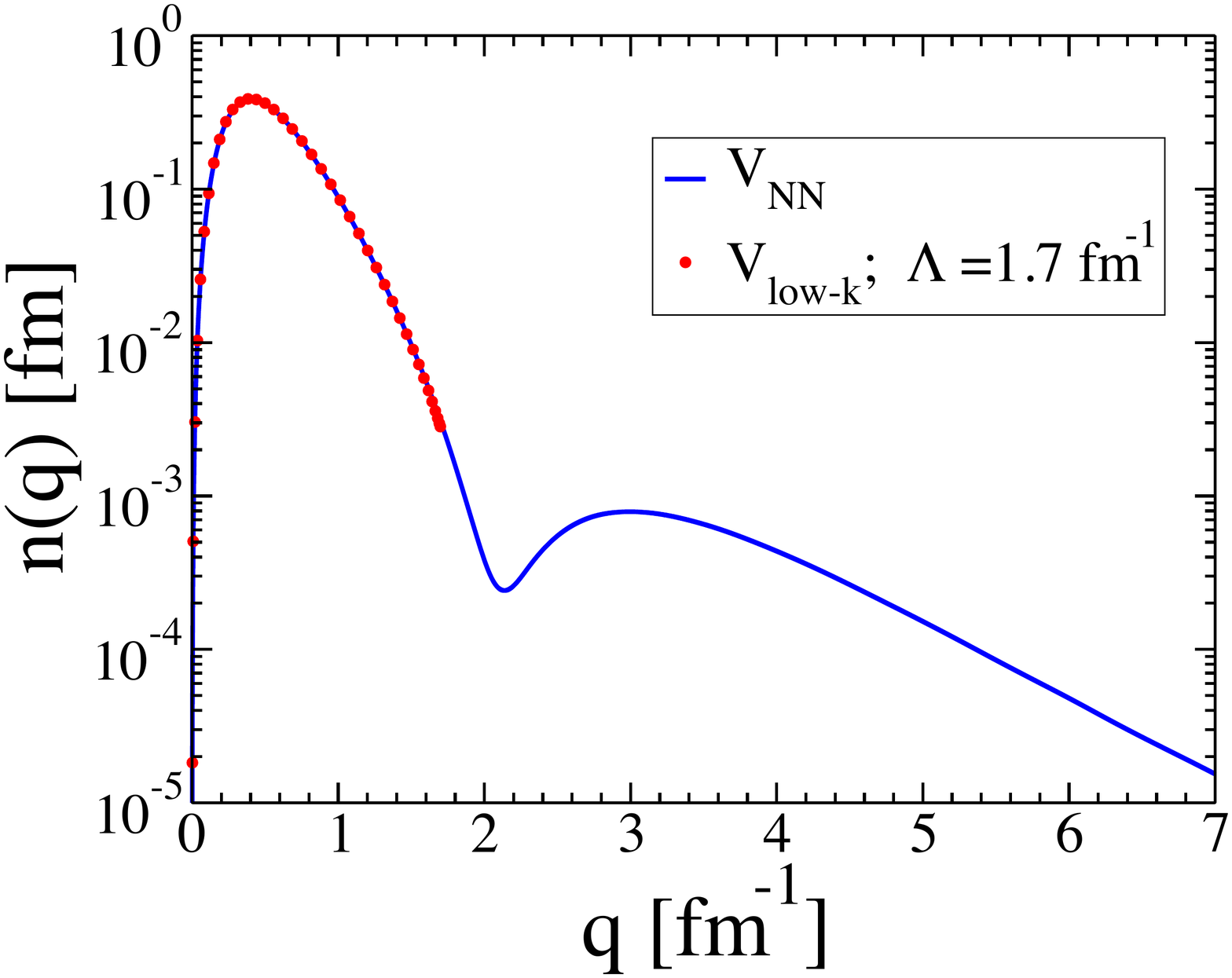}  &
\hspace{.cm} 
\includegraphics[width=2.5in]{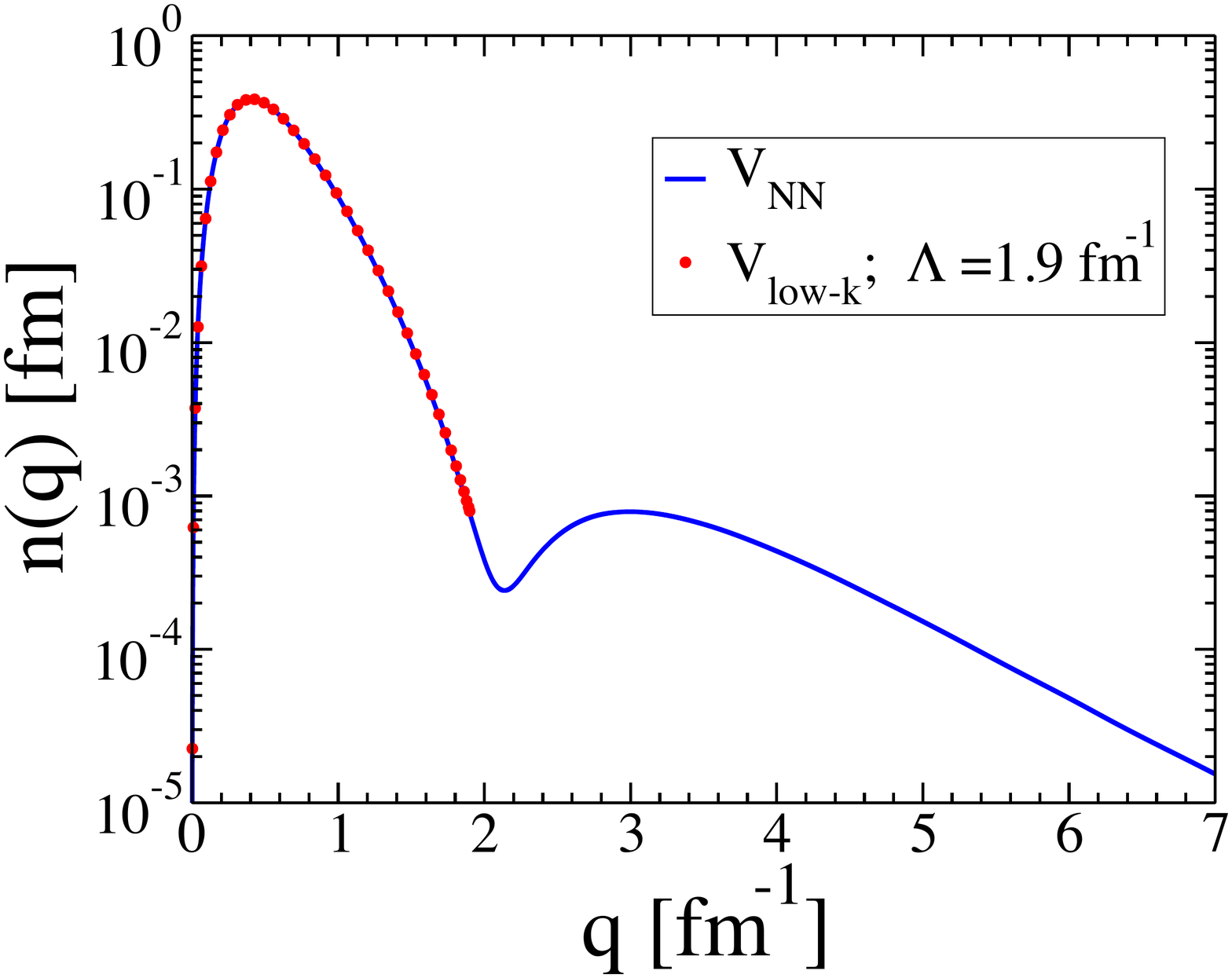}   \\   
\includegraphics[width=2.5in]{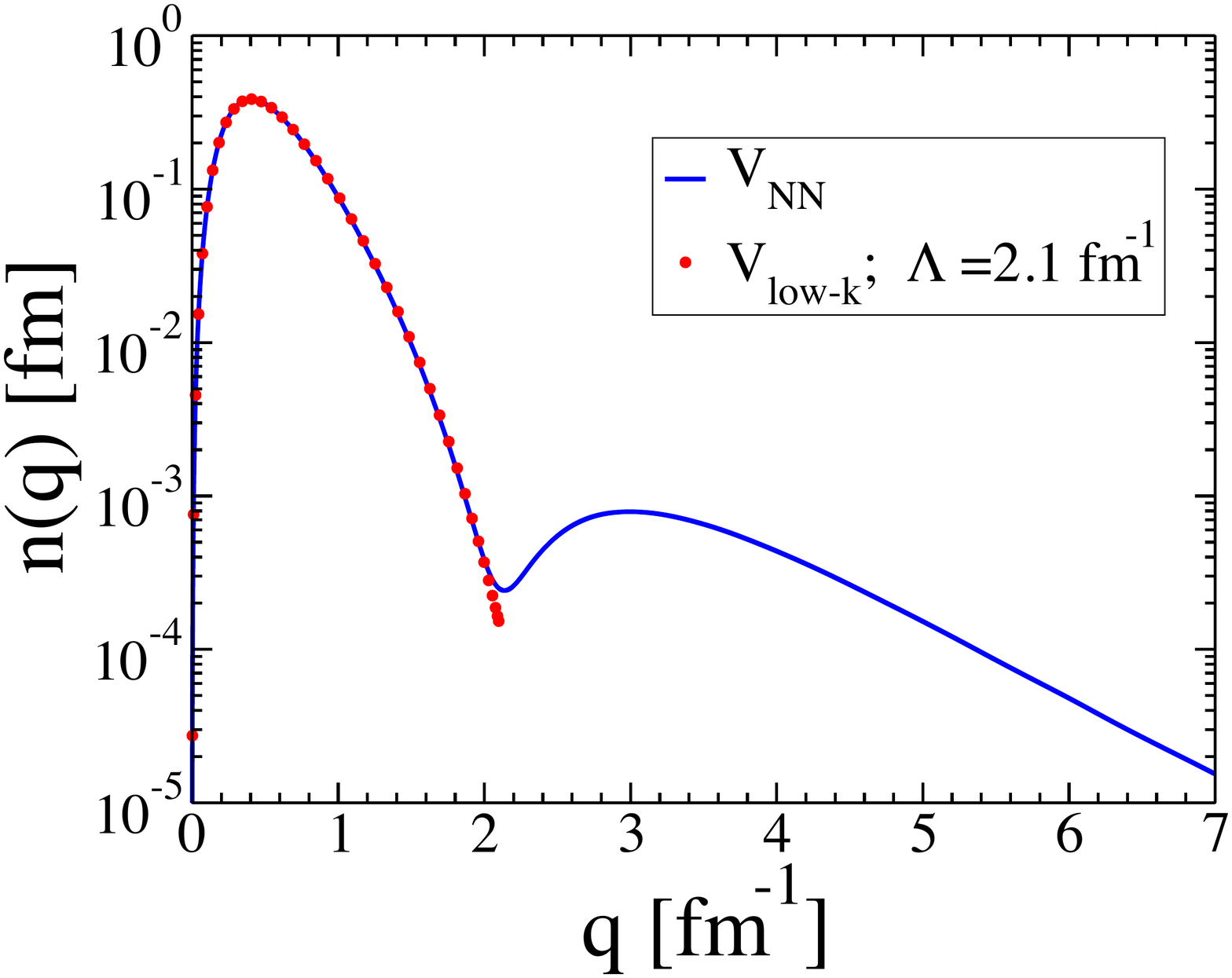} &
\hspace{.cm}
\includegraphics[width=2.5in]{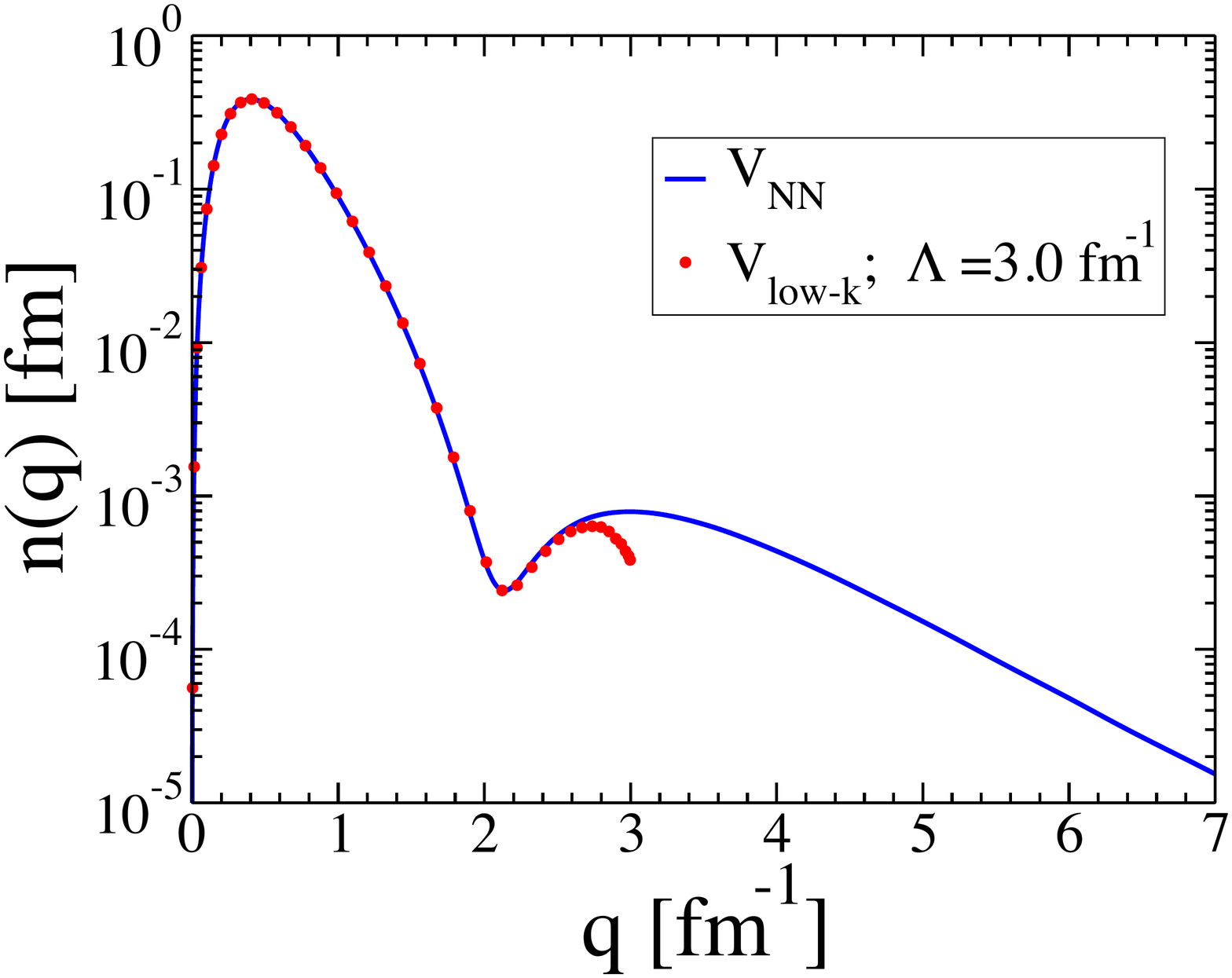}  \\
\includegraphics[width=2.5in]{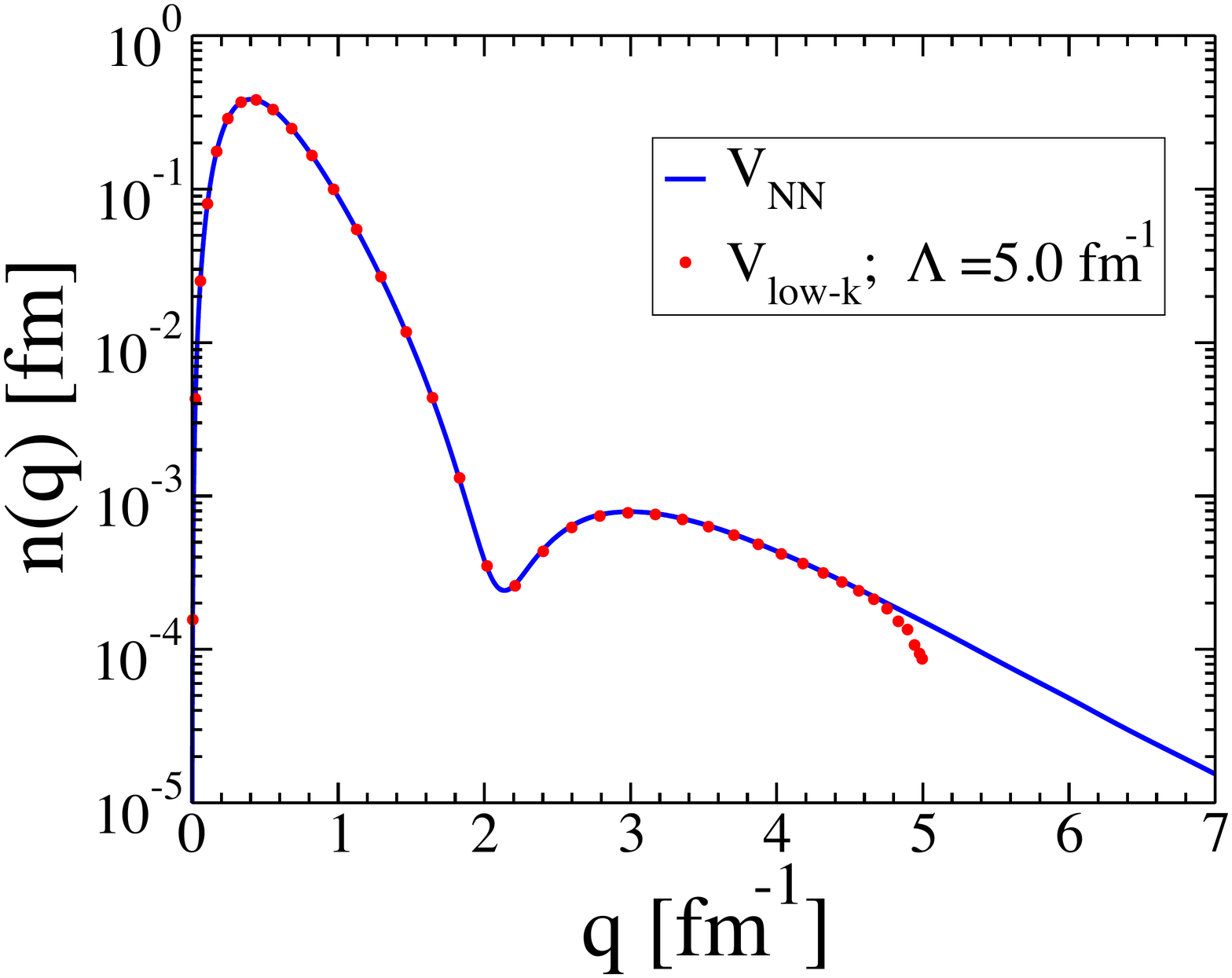}   &     
\hspace{.cm} 
\includegraphics[width=2.5in]{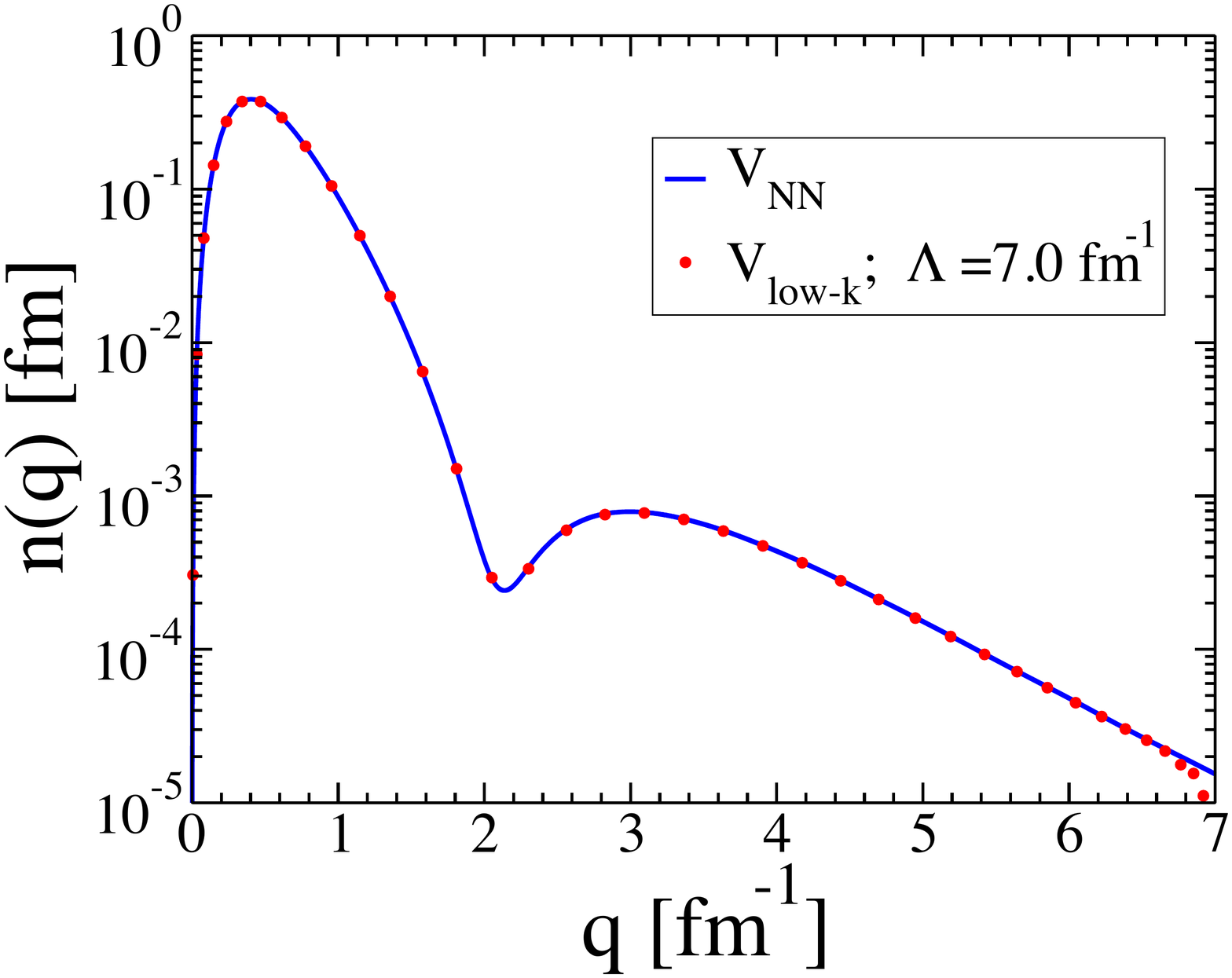} 
\end{array}
\]
\end{center}
\caption{ The momentum distribution function $n(q)$ calculated from MT-V bare and low momentum potential for $\Lambda=1.2-7.0\, fm^{-1}$}
\label{fig:nq}
\end{figure}

In order to simplify our analysis of the symmetrized 3B wave function $\Psi(p,q,x)$, which is calculated from Eq. (\ref{eq.WF-momentum}) after having the Faddeev component $\psi(p,q,x)$ from Eq. (\ref{eq.FC-magnitude}), and to obtain an insight on how the momentum is shared among the Jacobi coordinates, we have calculated the momentum probability densities which are defined as:
\begin{equation}
n(u_i)= 2 \pi u_i^2 \int_0^{\infty} du_j u_j^2 \int_{-1}^{+1} dx \,
\Psi^2(p,q,x) , \quad 4\pi \int_0^{\infty}  n(u_i) du_i = 1,
 \label{eq.Densities}
 \end{equation}
where $(i,j)=(1,2),\, (2,1)$, $u_1$ and $u_2$ stand for Jacobi momenta $p$ and $q$, respectively. The momentum probability densities $n(p)$ and $n(q)$ calculated for low momentum potentials for different values of cutoff $\Lambda$ are presented in Figs. \ref{fig:np} and \ref{fig:nq} and compared with corresponding results obtained from bare potential.
The momentum probability density $n(p)$ has a dip around $2\,fm^{-1}$, which has been shifted to down for $\Lambda$ less than $2.0\, fm^{-1}$ and for larger values it has been moved to up. Similar behavior can be seen in the momentum probability density $n(q)$, but the shift is not visible as much as $n(p)$.

\subsection{Expectation Values}\label{}
In order to test the accuracy of our numerical solution for Faddeev integral equation (\ref{eq.FC-magnitude}) and also calculation of total wave function (\ref{eq.WF-momentum}), we have calculated the expectation value of 3B Hamiltonian $\langle \Psi | H | \Psi \rangle \equiv \langle H \rangle$ and compared with calculated binding energy. The explicit form of the expression $\langle H \rangle$ which should be obtained from expectation values of free Hamiltonian $\langle H_0 \rangle$ as well as potential $\langle V \rangle$ are given in Ref. \cite{Elster-FBS27}. 
Our numerical results for $\langle H_0 \rangle$, $\langle V \rangle$ and consequently $\langle H \rangle$ calculated for low momentum interaction with $\Lambda=2.1\, fm^{-1}$ are given in Table \ref{table:Expectation}. We have done this test of numerical accuracy for different number of mesh points for Jacobi momenta $N_{jac}$ and angle variables $N_{sph}$, from 20 to 40. The comparison between the expectation value of 3B Hamiltonian $\langle H\rangle$ and eigenvalue energy $E_3$ shows that our results are in good agreement, however a better agreement can be reached if we consider a larger number of mesh points in our calculations.

\begin{table}[H]
\begin{tabular}{lccccccccccccccccl}
\hline \hline
  $N_{jac}$  && $N_{sph}$     && $\langle H_0 \rangle$ [MeV]  && $\langle V\rangle$ [MeV]   && $\langle H\rangle $ [MeV]  && $E_3$ [MeV] \\
  \hline
  20 && 20 &&  24.778  && -32.575 &&  -7.797 &&  -7.795 \\
  30 && 20 && 24.786  &&  -32.587 &&  -7.801 && -7.793 \\
  30 && 30 &&  24.786    && -32.587 && -7.801  &&  -7.793 \\
  40 && 40 &&    24.788 && -32.589  && -7.801  && -7.792  \\
\hline \hline
\end{tabular}
\caption {The expectation values $\langle H_0 \rangle$, $\langle V\rangle$ and $\langle H\rangle $ calculated for low momentum interaction constructed from MT-V potential for $\Lambda=2.1 \, fm^{-1}$.}
\label{table:Expectation}
\end{table}

\section{Summary}\label{Summary}

In this paper, we have performed 3B Faddeev calculations by employing a low-momentum potential $V_{low-k}$ which is derived in a non-PW representation and by using the Lee-Suzuki similarity transformation method. The main purpose of this work is to test the 3D form of low momentum interaction in three-body bound state calculations. 
The motivation of using 3D approach is avoiding the truncation problems and the necessity of complicated recoupling algebra that accompanies PW based calculations, instead the equations and amplitudes are formulated in the 3D approach, directly as a functions of momentum vector variables.
We have studied the dependence of three-body binding energy to low-momentum cutoff $\Lambda$, which separates the Hilbert space into a low momentum and a high momentum part, and have compared the results with those obtained by using bare spin-independent Malfliet-Tjon potential. 
The stability of our numerical results for low momentum interaction and 3B binding energy 
has been studied with the calculation of the expectation value of the total Hamiltonian
The agreement between binding energy results from low-momentum interaction and those obtained by bare one indicates that the 3D form of low momentum interaction is suitable to be applied in few-body calculation in a non PW representation.

After this first successful application of 3D form of $V_{low-k}$, there is now motivation to pursue further few-body bound and scattering calculations with the non PW form of low momentum potential and we predict that the incorporation of three-body force will be less cumbersome in a 3D approach.

\section*{Acknowledgment}

 We would like to thank Ch. Elster for useful discussions and to S. Bayegan and M. Harzchi for supplying the matrix elements of $V_{low-k}$. 
This work was supported in part by the Brazilian agency
Funda\c c\~ao de Amparo \`a Pesquisa do Estado de S\~ao Paulo.



\begin{thebibliography}{9}


\bibitem{Bogner-NPA684}
S. K. Bogner, T. T. S. Kuo and L. Coraggio, Nucl. Phys. A {\bf 684}, 432 (2001).

\bibitem{Bogner-PR386}
S. K. Bogner, T. T. S. Kuo and A. Schwenk, Phys. Rep. {\bf 386}, 1 (2003).

\bibitem{Bogner-PLB576}
S. K. Bogner, T. T. S. Kuo, A. Schwenk, D. R. Entem and R. Machleidt, Phys. Lett. B {\bf 576}, 265 (2003).


\bibitem{Lee-PLB91} 
S. Y. Lee and K. Suzuki, Phys. Lett. B {\bf91}, 173 (1980).

\bibitem{Suzuki-PTP92} 
K. Suzuki and R. Okamoto, Prog. Theor. Phys. {\bf 92}, 1045 (1994).


\bibitem{Bayegan-NPA814} 
S. Bayegan, M. Harzchi, M. R. Hadizadeh, Nucl. Phys. A {\bf 814}, 21 (2008). 

\bibitem{Bayegan-NPA832} 
S. Bayegan, M. Harzchi and M. A. Shalchi, Nucl. Phys.  A {\bf 832}, 1 (2010). 

\bibitem{Malfliet-NPA127}
R. A. Malfliet and J. A. Tjon, Nucl. Phys. A {\bf 127}, 161 (1969).

\bibitem{Fujii-PRC70}
S. Fujii et al., Phys. Rev. C {\bf 70}, 024003 (2004).

\bibitem{Nogga-PRC70}
Andreas Nogga, Scott K. Bogner, and Achim Schwenk, Phys. Rev. C {\bf 70}, 061002(R) (2004). 

\bibitem{Bogner-PRC75} 
S. K. Bogner, R. J. Furnstahl, and R. J. Perry, Phys. Rev. C {\bf 75}, 061001(R) (2007).

\bibitem{Bayegan-PRC79}  
S. Bayegan, M. A. Shalchi, M. R. Hadizadeh, Phys. Rev. C. {\bf 79}, 057001 (2009).

\bibitem{Bayegan-PRC77} 
S. Bayegan, M. R. Hadizadeh, and M. Harzchi, Phys. Rev.  C {\bf 77}, 064005 (2008). 

\bibitem{Bayegan-PTP120} 
S. Bayegan, M. R. Hadizadeh, and W. Gl\"{o}ckle, Prog. Theor. Phys. {\bf 120}, 887 (2008).

\bibitem{Hadizadeh-PRC83} 
M. R. Hadizadeh, L. Tomio, S. Bayegan, Phys. Rev. C {\bf 83}, 054004 (2011).

\bibitem{Elster-FBS24} 
Ch. Elster, J. H. Thomas and W. Gl\"{o}ckle, Few Body Syst. 24, 55 (1998).


\bibitem{Elster-FBS27}  
Ch. Elster, W. Schadow, A. Nogga and W. Gl\"ockle, Few-Body Syst. {\bf 27}, 83 (1999).

\bibitem{Hadizadeh-PRA85} 
M. R. Hadizadeh, M. T. Yamashita, Lauro Tomio, A. Delfino, T. Frederico, Phys. Rev. A {\bf 85}, 023610 (2012).

\bibitem{Hebeler-PRC85} Kai Hebeler, Phys. Rev. C {\bf 85}, 021002 (2012).

\end{thebibliography}
\end{document}